\documentclass[acmsmall]{acmart}

\usepackage{makecell}
\usepackage{subfigure}
\usepackage{bm}
\usepackage{enumitem}
\usepackage{multirow}
\usepackage{float}
\usepackage{tabularx}
\usepackage{pifont}
\newcommand{\xmark}{\ding{55}}

\usepackage{siunitx}
\AtBeginDocument{%
  \providecommand\BibTeX{{%
    \normalfont B\kern-0.5em{\scshape i\kern-0.25em b}\kern-0.8em\TeX}}}

\usepackage{xspace}
\newcommand\ie{i.\,e.\xspace}
\newcommand\eg{e.\,g.\xspace}

\newcommand\US{U.\,S.\xspace}

\setcopyright{rightsretained}
\acmJournal{PACMHCI}
\acmYear{2024} \acmVolume{8} \acmNumber{CSCW2} \acmArticle{428} \acmMonth{11}\acmDOI{10.1145/3686967}

\begin{document}

\title[Did the Roll-Out of Community Notes Reduce Engagement With Misinformation on X/Twitter?]{Did the Roll-Out of Community Notes Reduce Engagement With Misinformation on X/Twitter?}

\author{Yuwei Chuai}
\affiliation{
  \institution{University of Luxembourg}
  \country{Luxembourg}}
\email{yuwei.chuai@uni.lu}
\orcid{0000-0001-6181-7311}

\author{Haoye Tian}
\affiliation{
  \institution{University of Melbourne}
  \country{Australia}}
\email{haoye.tian@unimelb.edu.au}
\orcid{0000-0002-8049-3997}

\author{Nicolas Pr{\"o}llochs}
\affiliation{
 \institution{JLU Giessen}
 \country{Germany}}
\email{nicolas.proellochs@wi.jlug.de}
\orcid{0000-0002-1835-7302}

\author{Gabriele Lenzini}
\affiliation{
  \institution{University of Luxembourg}
  \country{Luxembourg}}
\email{gabriele.lenzini@uni.lu}
\orcid{0000-0001-8229-3270}


\begin{abstract}
Developing interventions that successfully reduce engagement with misinformation on social media is challenging. One intervention that has recently gained great attention is X/Twitter's Community Notes (previously known as ``Birdwatch''). Community Notes is a crowdsourced fact-checking approach that allows users to write textual notes to inform others about potentially misleading posts on X/Twitter. Yet, empirical evidence regarding its effectiveness in reducing engagement with misinformation on social media is missing. In this paper, we perform a large-scale empirical study to analyze whether the introduction of the Community Notes feature and its roll-out to users in the \US and around the world have reduced engagement with misinformation on X/Twitter in terms of retweet volume and likes. We employ Difference-in-Differences (DiD) models and Regression Discontinuity Design (RDD) to analyze a comprehensive dataset consisting of all fact-checking notes and corresponding source tweets since the launch of Community Notes in early 2021. Although we observe a significant increase in the volume of fact-checks carried out via Community Notes, particularly for tweets from verified users with many followers, we find no evidence that the introduction of Community Notes significantly reduced engagement with misleading tweets on X/Twitter. Rather, our findings suggest that Community Notes might be too slow to effectively reduce engagement with misinformation in the early (and most viral) stage of diffusion. Our work emphasizes the importance of evaluating fact-checking interventions in the field and offers important implications to enhance crowdsourced fact-checking strategies on social media.
\end{abstract}


\received{July 2023}
\received[revised]{January 2024}
\received[accepted]{March 2024}

\maketitle

\section{Introduction}

Misinformation refers to false or misleading information that is contradicted by empirical evidence or is inconsistent with common or expert consensus \cite{ecker_psychological_2022,wu_misinformation_2019,freiling_believing_2021}. The spread of misinformation has become a pervasive problem in the digital age, particularly on popular social media platforms such as X (formerly known as Twitter), TikTok, and Facebook. Online engagement with misinformation poses significant threats to the stability of society and contributes to contentious events across various domains, including, but not limited to, politics, finance, and public health \cite{ecker_psychological_2022,chuai_anger_2022,gradon_countering_2021}. During political elections, misinformation can influence voter opinions and behaviors, potentially undermining the democratic process \cite{allcott_social_2017, mathur_how_2022, suciu_bots_2022}. In financial markets, misinformation can lead to market volatility and financial losses for investors and businesses. For example, 130\$ billion in stock value was wiped out in minutes following a false tweet indicating that an ``explosion'' injured Barack Obama in 2013 \cite{rapoza_can_2017}. Online misinformation can even pose severe risks to public health. For instance, during the COVID-19 pandemic, misinformation contributed to individuals' hesitancy and reluctance to accept vaccinations, resulting in a heightened risk of infection \cite{enders_relationship_2022,loomba_measuring_2021}. Consequently, mitigating user engagement with online misinformation is crucial and urgent.

To reduce engagement with online misinformation, various interventions have been proposed \cite{he_reinforcement_2023,ecker_psychological_2022}. These interventions can be broadly divided into two categories, namely, prebunking and debunking \cite{ecker_psychological_2022,wu_misinformation_2019,he_reinforcement_2023,saltz_misinformation_2021}. Prebunking indicates pre-emptive interventions that aim to help individuals identify and resist misinformation they may subsequently encounter \cite{ecker_psychological_2022}. For example, a popular strategy is redirection, which focuses on directing users' attention towards trustworthy information sources \cite{ecker_psychological_2022,saltz_misinformation_2021}. While prebunking interventions can assist individuals in developing resistance to misinformation in a relatively broad context, they remain inadequate to counteract specific instances of misinformation that can quickly become viral and cause concern \cite{ecker_psychological_2022}. The other category of interventions, debunking, aims to fact-check the specific misinformation after its creation. Debunking is frequently carried out by third-party organizations, such as Snopes and PolitiFact, which recruit professional fact-checkers and subject experts to ensure the reliability of the verdicts. However, due to the limited number of human experts, it's difficult to implement professional fact-checking in a timely and scaled manner. As an alternative, automatic fact-checking based on machine learning models can process large numbers of posts at a high speed. However, machine learning typically shows low performance in fact-checking real-world social media content due to the lack of counter-evidence in training data \cite{glockner_missing_2022}. 

In order to address the drawbacks of the existing approaches to debunking, recent research has proposed to employ non-expert fact-checkers in the crowd to fact-check social media posts at scale \cite{allen_scaling_2021,saeed_crowdsourced_2022}. Though individual assessments can have bias and noise, the aggregated judgment of even relatively small crowds has been found to be comparable to those from expert fact-checkers \cite{epstein_will_2020,bhuiyan_investigating_2020,drolsbach_diffusion_2023}. Informed by these research findings, X/Twitter has launched its Community Notes platform (previously known as ``Birdwatch''), which is the first large-scale effort of crowdsourced fact-checking deployed in practice \cite{wojcik_birdwatch_2022, saeed_crowdsourced_2022}. The Community Notes platform allows users to identify tweets they believe are misleading or not misleading and write textual notes that provide context to the tweet. The Community Notes platform also features a rating system that allows users to rate the helpfulness of the fact-checks. If a note receives a high helpfulness score, it is displayed directly on the fact-checked tweets, visible to all users on X/Twitter. In an attempt to foster transparency, X/Twitter makes all Community Notes contributions available for analysis by researchers \cite{prollochs_community-based_2022}. This offers a unique opportunity to analyze how users interact with this misinformation intervention in a real-world setting. 

While the Community Notes feature has garnered considerable attention lately, an understanding of whether it is effective in reducing engagement (\eg, retweet volume, likes) with misinformation on X/Twitter is missing. Previous research has shown that the fact-checks carried out on Community Notes are relatively accurate and trustworthy \cite{saeed_crowdsourced_2022,drolsbach_diffusion_2023,drolsbach_community_2024}, and that Community Notes contributors focus on content posted by individuals with whom they hold opposing political views \cite{allen_birds_2022}. Moreover, during the pilot phase of the Community Notes feature, X/Twitter conducted survey experiments to show that note annotations were, to a large extent, perceived as helpful by other users \cite{wojcik_birdwatch_2022}. Users exposed to notes were also less likely to retweet or like misleading posts \cite{wojcik_birdwatch_2022}. Additionally, recent research has analyzed the retweet volume (\ie, the engagement in sharing) of posts that were fact-checked via Community Notes \cite{drolsbach_diffusion_2023}. Contrary to misinformation fact-checked by third-party fact-checking organizations \cite{vosoughi_spread_2018}, community fact-checked misinformation received fewer retweets than truthful posts during the pilot phase of Community Notes, which can be partially attributed to the fact that community contributors were more likely to fact-check high-influence accounts with a large number of followers \cite{drolsbach_diffusion_2023}. However, the key question of whether Community Notes is an effective intervention to reduce engagement with misinformation on social media has, so far, not been studied. This is the objective of this work.

In this paper, we conduct a comprehensive empirical study to investigate the effect of the Community Notes feature on user engagement with misinformation on X/Twitter in terms of retweet volume and likes. Our study covers all notes and associated tweets posted on X/Twitter between July 1, 2021, and April 6, 2023. This observation period includes two important intervention dates: October 6, 2022, and December 11, 2022. The first intervention occurred on October 6, 2022, when Community Note's pilot phase ended and helpful notes became visible to users in the \US \cite{twitter_helpful_2022}. The second intervention occurred on December 11, 2022, when X/Twitter expanded Community Notes program to the rest of the world. This longitudinal dataset provides us with a unique opportunity to examine the impact of the roll-out of Community Notes on the engagement with misinformation on X/Twitter. In our analysis, we first examine the frequency of notes per tweet and the time it takes for notes to be created and displayed to users. Second, we use the tweets posted before and after the interventions to longitudinally examine the treatment effect with Difference-in-Differences (DiD) models at the feature roll-out level. This approach enables us to gauge the relative reduction in engagement with misinformation after the roll-out of the Community Notes feature compared to the pre-roll-out period. Third, we conduct a placebo analysis to validate the results obtained from the DiD models. Fourth, we limit our analysis to the tweets posted after the interventions and use Regression Discontinuity Design (RDD) as a complementary method to test the discontinuity (\ie, treatment effect) at the note display level. This approach allows us to measure the relative reduction in engagement with misinformation on which notes are displayed compared to misinformation on which notes are not displayed. Finally, we conduct a comprehensive robustness analysis and consider potential confounding factors from multiple levels to further consolidate the findings of this study.

Our study finds no evidence to suggest that the roll-out of the Community Notes feature to users in the \US and around the world reduced engagement with misinformation on X/Twitter in terms of retweet count and like count. This finding is confirmed by both DiD and RDD models. Although we observe a significant increase in the volume and speed of fact-checking, particularly for tweets from verified users with many followers, our results suggest that the response time of Community Notes, \ie, from the time of tweet creation to the time of displaying notes, is still insufficient to reduce engagement with misinformation in the early (and most viral) stage of diffusion. Additionally, we observe a decline in engagement with fact-checked tweets -- irrespective of the veracity -- after the expansion of the Community Notes feature to a global audience. Interestingly, this decline disappears when we exclusively analyze the tweets from the set of authors that have been fact-checked both before and after the expansion. This suggests that the decrease in engagement with community fact-checked tweets can be partially attributed to differences in the fact-checking targets. Altogether, our results highlight the importance of evaluating misinformation interventions within real social media environments. We provide an in-depth discussion on the theoretical and practical implications of our work, which furnishes valuable insights into improving crowdsourced fact-checking strategies on social media.

\section{Background and Related Work}
\subsection{Misinformation on Social Media}

Misinformation is generally defined as false or misleading information that is contradicted by empirical evidence or is inconsistent with common or expert consensus \cite{ecker_psychological_2022,wu_misinformation_2019,freiling_believing_2021}.\footnote{In previous research, a variety of terms have been used to refer to false and misleading information. Examples include ``misinformation,'' ``disinformation,'' and ``fake news,''  each of which places a different emphasis on the underlying characteristics. Disinformation indicates a subset of misinformation that is spread with the intention of harm or deception \cite{wu_misinformation_2019,freiling_believing_2021}. Notably, detecting intentions, especially those not explicitly expressed in the content, can be challenging \cite{krause_infodemic_2022}. Fake news is a type of misinformation in the form of news that is often fabricated to mislead people, especially in the political and financial fields \cite{wu_misinformation_2019,chuai_anger_2022,shu_fake_2017}.
In our work, we use the term ``misinformation'' as an umbrella as it does not impose assumptions on the motif or the format of the content  \cite{ecker_psychological_2022,wu_misinformation_2019,freiling_believing_2021}.} When people believe and give credibility to misinformation, the notion of truth becomes corrupted, leading to potentially harmful consequences. This is particularly problematic in the digital age, where users are exposed to persuasive online information with uncertain origins and a lack of authenticity \cite{perez-escolar_systematic_2023}. 

The proliferation of social media as the primary source of information has led to a rise in online misinformation with the lack of a ``gate-keeper'' (\eg, professional editors), which poses a significant threat to the stability of society and contributes to contentious events across different domains such as, for example, politics, finance, and public health \cite{ecker_psychological_2022,bak-coleman_combining_2022}. In contemporary democratic politics, social media has emerged as the main platform for political propaganda. Dedicated misinformation designed to pursue political agendas easily gets viral engagement online \cite{druckman_threats_2022,vosoughi_spread_2018,papakyriakopoulos_impact_2022}. During political elections, people are exposed to online misinformation more often than usual, which could influence individuals' opinions and voting behaviors, and even undermine elections \cite{mathur_how_2022,mosleh_measuring_2022,suciu_bots_2022,recuero_hyperpartisanship_2020,haque_combating_2020}. For example, during the 2016 \US presidential election, on average, each \US adult saw and remembered one or perhaps several false stories with clear partisan leanings on social media \cite{allcott_social_2017}. Similar problems also arose in the Global South, for example, during the 2018 Brazilian presidential election \cite{haque_combating_2020,recuero_hyperpartisanship_2020}. Politics and finance are closely linked, and political false information can accordingly incite shocks in the financial market. For instance, in 2013, \$130 billion in stock value was wiped out in a few minutes following a false tweet indicating that an ``explosion'' injured Barack Obama \cite{rapoza_can_2017}. Additionally, false information about companies or economic indicators can also lead to stock market volatility, potentially resulting in financial losses for investors and businesses \cite{clarke_fake_2021}. Online misinformation can even pose severe risks to public health. For example, the COVID-19 pandemic has been accompanied by an ``infodemic'' around the world, characterized by the dissemination of a plethora of online misinformation regarding the virus, its origin, and potential treatments or prevention strategies \cite{briand_infodemics_2021}. A global survey demonstrated that a substantial proportion of people view COVID-19 misinformation as reliable \cite{roozenbeek_susceptibility_2020}. Beliefs in misinformation during COVID-19 influenced people's hesitancy and refusal of getting vaccinated, which affected their risk of infection \cite{enders_relationship_2022,pierri_online_2022,sultana_dissemination_2021,akbar_misinformation_2021,varanasi_accost_2022,roozenbeek_susceptibility_2020}.

Given the detrimental effects of misinformation across society and around the world, user engagement with online misinformation has become a pressing issue that social media platforms, policymakers, and researchers continue to grapple with \cite{lazer_science_2018}.

\subsection{Misinformation Interventions}
As concerns regarding the impact of online misinformation on online users continue to grow, many intervention strategies have been proposed to reduce engagement with misinformation and ensure that people remain accurately informed. These interventions can be broadly distinguished into two categories, namely, prebunking and debunking \cite{ecker_psychological_2022}.

Prebunking is a kind of pre-emptive intervention and seeks to help people recognize and resist subsequently encountered misinformation \cite{ecker_psychological_2022,wu_misinformation_2019}. Prebunking interventions include factually correct information presentation, pre-emptive correction, digital media literacy training, and generic misinformation warning before the misinformation. Redirection, which focuses on shifting users' attention to trustworthy information sources, is a kind of prebunking intervention implemented in practice by social media platforms \cite{pennycook_accuracy_2022}. The prebunking approach provides a promising tool to act pre-emptively and help people build immunity to misinformation in a relatively general manner \cite{ecker_psychological_2022, pennycook_fighting_2020,pennycook_shifting_2021}. While prebunking interventions can assist individuals in developing resistance to misinformation in a relatively broad context, they remain inadequate to counteract specific instances of misinformation that can quickly become viral and cause concern \cite{ecker_psychological_2022}. 

Debunking, also known as fact-checking, involves responding to specific instances of misinformation after they have been exposed to demonstrate why they are false \cite{wu_misinformation_2019,ecker_psychological_2022,he_reinforcement_2023}. Reliable fact-checking plays a crucial role in reducing engagement with false or misleading information and promoting accurate information that can inform public opinion and decision-making. Fact-checking strategies can be implemented in various ways, including third-party organizations and automated models \cite{he_reinforcement_2023}. Of note, manual professional fact-checking carried out by experts is a time-consuming process. Verdicts are usually reached after a message has gone viral, and interventions can have only limited effect \cite{stein_realtime_2023}. Moreover, this strategy is difficult to scale due to the limited number of professional fact-checkers. To realize large-scale fact-checking and reduce response time, automatic fake news detection becomes a promising direction. Many machine learning models are trained based on the features of content (\eg, writing style) and context (\eg, network structure) to detect misinformation automatically. Nevertheless, because training data often lacks counter-evidence, fully automated fact-checking remains unrealistic \cite{glockner_missing_2022}.

A novel approach that has gained attention recently is to leverage the collective intelligence of the crowd, \ie, wisdom of crowd, by allowing individuals who receive online information to do the fact-checking at scale \cite{stein_realtime_2023,saeed_crowdsourced_2022,micallef_true_2022,bhuiyan_investigating_2020,allen_scaling_2021,yasseri_can_2023}. This approach is based on the expectation that non-expert fact-checkers can achieve comparable accuracy to that of experts through collective assessments \cite{drolsbach_diffusion_2023}. While individual assessments of non-expert fact-checkers can be unreliable and inconsistent, harnessing the collective judgment of a crowd can be an effective method for identifying misinformation. Recent studies demonstrate that even small crowds can produce accuracy levels comparable to those of expert fact-checkers \cite{epstein_will_2020,bhuiyan_investigating_2020,lazer_science_2018}.

In practice, various countermeasures against misinformation have been implemented by news media and social media platforms. Based on the redirection strategy, TikTok implemented a feature where a banner with links to trustworthy information sources was attached to COVID-19 related videos \cite{tiktok_taking_2019}. Google Search and YouTube have implemented a knowledge panel feature, which provides users with high-level credible information related to their search terms \cite{google_fighting_2019}. Instagram and Facebook apply labels to posts based on third-party fact-checker ratings to provide users with more accurate information \cite{meta_metas_2023}. For example, Snopes and PolitiFact are two pioneering \US-based third-party fact-checking organizations. They typically employ trained journalists or subject matter experts to investigate and verify the accuracy of claims made in news articles, social media posts, and other forms of online content \cite{micallef_true_2022}. Also, the International Fact-Checking Network (IFCN) at Poynter Institute was launched to bring together fact-checkers from more than 100 organizations around the world and support the global fight against misinformation \cite{haque_combating_2020}. So far, however, such large-scale fact-checking efforts (and research) are mainly focused on the Global North. In the Global South, fact-checking efforts are still in its infancy and face additional challenges (\eg, due to under-resourced languages, lack of fact-checking sources) \cite{juneja_human_2022,haque_combating_2020}. Ultimately, X/Twitter recently implemented the Community Notes feature as a significant step to combat misinformation on their platform \cite{twitter_community_2023}. This innovative feature marks the first large-scale effort of crowdsourced fact-checking deployed in practice \cite{saeed_crowdsourced_2022}. 

Altogether, various misinformation interventions have been proposed by researchers or implemented by social media platforms \cite{ecker_psychological_2022,pennycook_fighting_2020,saeed_crowdsourced_2022,tiktok_taking_2019,google_fighting_2019,meta_metas_2023}. Yet, their effectiveness in reducing engagement with misinformation is oftentimes unclear. It is thus essential to evaluate the effectiveness of misinformation interventions directly on social media platforms.

\subsection{Community Notes on X/Twitter}
Community Notes is a feature implemented on X/Twitter that allows registered users, known as contributors, to independently fact-check tweets and add context in the form of notes \cite{wojcik_birdwatch_2022,prollochs_community-based_2022}. Contributors can label tweets as ``misinformed or potentially misleading'' (misinformation) or ``not misleading'' (true information). The statuses of notes are determined by an algorithm that calculates the helpfulness score based on the ratings made by the contributors from different perspectives \cite{wojcik_birdwatch_2022}. Only notes that mark tweets as ``misinformed or potentially misleading'' and receive high helpfulness scores earn the status of ``Helpful'' and are displayed on the corresponding tweets. X/Twitter launched the pilot program of Community Notes on January 28, 2021, and limited its visibility to a small number of pilot participants. Thus, during the pilot phase, the community fact-checks were not expected to directly influence engagement with the fact-checked tweets \cite{drolsbach_diffusion_2023}. The Community Notes feature was later expanded to all users in the \US on October 6, 2022, and to global users on December 11, 2022 \cite{twitter_helpful_2022}.

In an attempt to foster transparency, X/Twitter makes all Community Notes contributions available for analysis by researchers \cite{prollochs_community-based_2022}. Prior works have used the data to study the mechanisms and behaviors of raters on the platform \cite{prollochs_community-based_2022,allen_birds_2022}. Note contributors write more notes for misleading than not misleading tweets \cite{prollochs_community-based_2022}, tend to focus on content posted by individuals with whom they hold opposing political views \cite{allen_birds_2022}, and are more likely to fact-check posts from influential user accounts with many followers \cite{drolsbach_diffusion_2023}. Furthermore, research has shown that fact-checks for tweets from influential accounts yield a lower level of consensus among the contributors \cite{prollochs_community-based_2022} and that the assessments made on Community Notes are comparable to those obtained from human experts \cite{saeed_crowdsourced_2022}. Recent research has also used the fact-checked tweets on Community Notes to investigate the diffusion patterns of misinformation on X/Twitter \cite{drolsbach_believability_2023,drolsbach_diffusion_2023}. Contrary to misinformation fact-checked by third-party fact-checking organizations \cite{vosoughi_spread_2018}, community fact-checked misinformation received fewer retweets than truthful posts during the pilot phase of Community Notes \cite{drolsbach_diffusion_2023}. Additionally, during the pilot phase, a test conducted by X/Twitter demonstrated that Community Notes could reduce users' intentions to share misinformation \cite{wojcik_birdwatch_2022}. 

In sum, while the Community Notes feature has garnered considerable attention lately, it still remains unclear whether it is an effective tool in reducing engagement with misinformation among actual users on the social media platform. The half-life of a tweet is short: about 95\% of tweets have no relevant impressions after two days, and it only takes about 79.5 minutes before half of impressions are created for a tweet \cite{pfeffer_half-life_2023}. Hence, even if Community Notes reduces sharing intentions, the time delay until a community note becomes visible may be too long to effectively reduce engagement with misinformation in the early (and most viral) stage of the diffusion. Further research is needed to determine the actual impact of Community Notes on the overall engagement with misinformation on X/Twitter.

\section{Materials and Methods}
To evaluate the effectiveness of Community Notes in reducing engagement with misinformation on X/Twitter, we first download all note contributions on the Community Notes platform. Second, through X/Twitter's Academic Research API, we collect the fact-checked tweets using the tweet IDs referenced in Community Notes. Third, we measure engagement using retweet and like counts, and extract a wide variety of additional variables that may affect the engagement with content on X/Twitter (\eg, the number of followers, emotions, topics) from both notes and tweets. Subsequently, we employ Difference-in-Differences (DiD) as the primary method to analyze whether the introduction of the Community Notes feature and its roll-out to users in the \US and around the world has reduced engagement with misinformation on X/Twitter (\ie, at the feature roll-out level). Additionally, we utilize Regression Discontinuity Design (RDD) as a complementary method to examine the effectiveness of Community Notes at the note display level. The latter approach allows us to measure the relative reduction in engagement with misleading posts on which notes are displayed compared to those on which notes are not displayed.

\subsection{Data Collection}
\label{sec:data_collection}
All note contributions made on Community Notes are available for the public via a dedicated website.\footnote{\url{https://twitter.com/i/communitynotes/download-data}} The data used in our study was downloaded on April 8, 2023, and it covers all contributions up until April 6, 2023, such as notes, ratings, note status history, and user enrollment status. Specifically, the data includes 68,235 notes, 75,508 note status histories, 2,293,934 ratings, and 64,359 enrolled users. Notes deleted by users still retain their status histories and ratings. Each note contains its ID, referenced tweet ID, creation time, and classification (\ie, ``misinformed or potentially misleading'' or ``not misleading''). Each note status item records the change of referenced note statuses, including ``Needs More Ratings (NMR)'', ``Currently Rated Helpful (CRH)'', and ``Currently Rated Not Helpful (CRNH)'', which are determined by their helpfulness scores. All fact-checking notes on Community Notes start with the ``Needs More Ratings'' status until they receive at least 5 total ratings. To prevent one-side ratings and identify notes that are helpful to a wide range of people, Community Notes uses matrix factorization on the note-rater matrix to calculate note helpfulness scores. Notes marking tweets as misinformation (misinformed or potentially misleading) with a note helpfulness score of 0.40 and above earn the status of Helpful \cite{wojcik_birdwatch_2022}.

In addition to the contribution data, the note ranking algorithm code that X/Twitter runs in production is also released on GitHub.\footnote{\url{https://github.com/twitter/communitynotes/tree/main/sourcecode}} We download the open-source code to calculate the helpfulness score for each note. As of April 13, 2023, the source code is in version 1.0 and uses a multi-model note ranking strategy. Three note ranking models, namely, Core, Expansion, and Coverage, run the same ranking algorithm but differ in their ranking areas or thresholds. The Core model is always the most authoritative one. After running the code, we get the note score for each note. For the non-deleted notes, 63,870 out of 68,235 (93.6\%) are determined by the Core model.

After collecting the Community Notes data, we identify the tweets referenced by the notes based on the associated 45,664 unique tweet IDs. Utilizing the X/Twitter Academic Research API, we map these tweet IDs to the corresponding source tweets and successfully retrieve 39,268 tweets, which corresponds to 86\% of all tweets. Thus, the vast majority of the source tweets fact-checked via Community Notes are captured in our dataset. To address the concern that some potentially harmful misleading tweets with CRH notes may be removed by X/Twitter and cannot be fetched via IDs through the API \cite{jahanbakhsh_exploring_2023}, we specifically examine the subset of tweets with CRH notes. Out of 4,686 tweets with CRH notes, 3,872 (82.6\%) are successfully retrieved. This retrieval rate for tweets with CRH notes is similar to the overall rate for all tweets, which indicates a consistent level of data retrieval across categories. Furthermore, it is important to emphasize that X/Twitter explicitly states that notes that are rated helpful appear directly on the source tweets and have no effect on the display of source tweets or enforcement of X/Twitter's rules.\footnote{\url{https://communitynotes.twitter.com/guide/en/about/faq}}

We perform two filtering steps for our analysis. First, this study focuses on tweets posted in English and we filter out all source tweets that were not posted in the English language. This results in a total of 36,077 tweets authored by 15,348 unique users. Second, since Community Notes is a new feature introduced in January 2021 and requires time to gain contributors, we restrict our analysis to tweets posted after July 1, 2021. This allows for a period of six months for Community Notes to gain attention and for enough high-quality notes to be written and rated by the community. After filtering, we obtain 30,267 unique tweets posted between July 2021 and April 2023, which account for 83.9\% of the original set of English tweets. These tweets serve as the basis for our analysis and are subsequently divided into multiple groups to study the effect of Community Notes on reducing engagement with misinformation on X/Twitter.

\subsection{Variable Extraction}
\label{sec:variable_extraction}

In this paper, we examine the effect of the roll-out of Community Notes on the engagement with misinformation on X/Twitter using two commonly-used engagement metrics, namely (1) the retweet count and (2) the like count. Retweet count ($\mathit{RetweetCount}$) is a count variable denoting the number of retweets (not including quote tweets) a single tweet receives on X/Twitter. Tweets can reach a large number of people through retweets. Hence retweet count is the most used metric to measure the virality of posts on X/Twitter \cite{vosoughi_spread_2018,chuai_anger_2022,drolsbach_diffusion_2023}. Additionally, like count ($\mathit{LikeCount}$) is a count variable that measures how many users on X/Twitter have indicated that they appreciate or agree with the content of a tweet \cite{wojcik_birdwatch_2022}. The variables of $\mathit{RetweetCount}$ and $\mathit{LikeCount}$ are directly collected from the fields of public metrics embedded in the tweet objects.

Moreover, we use the X/Twitter Academic Research API to collect a wide variety of variables that may affect engagement with tweets on X/Twitter:
\begin{itemize}[leftmargin=*]
    \item $\bm{\mathit{WordNum}}$: A count variable indicating the number of words in a single tweet. Text length can have a positive effect on the perceived text quality \cite{chuai_anger_2022,ma_characterizing_2023}. To calculate the text length, we remove all links, stopwords, and non-alphabetic characters from the tweet text. We also remove the ``\#'' prefix from hashtags and retain the text. Additionally, we remove the ``@'' prefix from user accounts mentioned in the content, treating the usernames the same as the names of individuals mentioned in the content. $\mathit{WordNum}$ then corresponds to the number of remaining words in each tweet.
    \item $\bm{\mathit{Media}}$: A dummy variable indicating whether the tweet contains media elements, such as pictures or videos ($=$ 1; otherwise 0). Media elements can enhance user engagement with the corresponding information \cite{zhou_linguistic_2021}.
    \item $\bm{\mathit{Hashtag}}$: A dummy variable indicating whether the tweet contains one or more hashtags ($=$ 1; otherwise 0). By using one or more hashtags, users can link their tweets to larger conversations or hot topics. This can increase the visibility of their tweets \cite{chuai_what_2022}.
    \item $\bm{\mathit{Mention}}$: A dummy variable indicating whether the tweet contains one or more mention symbols ($=$ 1; otherwise 0). Users can mention other users in the content of tweets, and this can also affect engagement with the tweets \cite{chuai_what_2022}.
    \item $\bm{\mathit{AcctAge}}$: A count variable indicating the age of the account (in days) at the time of data collection.
    \item $\bm{\mathit{Verified}}$: A dummy variable indicating whether the author is verified ($=$ 1; otherwise 0).
    \item $\bm{\mathit{Followers}}$: A count variable indicating the number of followers of the author.
    \item $\bm{\mathit{Followees}}$: A count variable indicating the number of followees of the author.
\end{itemize}
In addition to the above variables, engagement may also be affected by emotions and topics in the fact-checked tweets \cite{chuai_anger_2022,vosoughi_spread_2018,robertson_negativity_2023,brady_emotion_2017,solovev_moral_2022,prollochs_emotions_2021}:

\begin{itemize}[leftmargin=*]
\item \textbf{Emotions}: To control the possible effects of emotions on the engagement with information, we consider six fine-grained basic emotions, namely, anger, disgust, joy, sadness, fear, and surprise \cite{chuai_anger_2022,vosoughi_spread_2018}. We first perform standard preprocessing steps (see $\mathit{WordNum}$ variable). Subsequently, we use Spacy, an Industrial-Strength Natural Language Processing library in Python, to lemmatize the words. Next, we count the frequency of words that belong to each of the emotions in each tweet based on the commonly used NRC emotion lexicon \cite{mohammad_crowdsourcing_2013}. As a result, 81.2\% of all tweets contain at least one emotion word. We then calculate the weight of each emotion in each tweet according to the number of corresponding emotion words. Notably, tweets without any emotion words are still included in the analysis, with the weights of all emotions being 0.

\item \textbf{Topics}: X/Twitter has already labeled the corresponding domains that each tweet belongs to in the collected dataset. We extract all the domains in the fact-checked tweets and classify them into six general topics, \ie, Entertainment, Business \& Finance, Politics, Science \& Technology, Emergency Events, and Ongoing News Story. The topic list is shown in Table \ref{tab:domains2topics} in the Appendix. For each topic, we have a corresponding dummy variable to denote whether the tweet belongs to it ($=$ 1; otherwise 0). Each tweet can have multiple topics. Thus, the sum of the topic dummy variables for each tweet is not constrained to equal one. 
\end{itemize}
Descriptive statistics for the extracted variables are reported in Table \ref{tab:desc_vars} in the Appendix.

\subsection{Empirical Models}

We evaluate the effectiveness of Community Notes using two popular quasi-experimental methods. First, we use tweets posted before and after the interventions to longitudinally examine the treatment effect with \emph{Difference-in-Differences (DiD)} models at the feature roll-out level. This approach enables us to gauge the additional reduction in engagement with misinformation on X/Twitter after the roll-out of Community Notes feature compared to the pre-roll-out period. Second, we limit our analysis to tweets posted after the interventions, and use \emph{Regression Discontinuity Design (RDD)} as a complementary method to test the discontinuity (\ie, treatment effect) at the note display level. Here, the treatment effect indicates the extra reduction in engagement with misinformation on which notes are displayed compared to the engagement with misinformation on which notes are not displayed.

\subsubsection{Difference-in-Differences (DiD) model}

We use Difference-in-Differences (DiD, ref. \cite{angrist_mostly_2009}) to estimate the treatment effect of Community Notes in reducing engagement with misinformation on X/Twitter at the feature roll-out level. Longitudinally, the DiD model assumes that the treatment group (the group affected by the treatment) and control group (the group not affected by the treatment) follow parallel trends in the absence of the intervention.\footnote{To ensure a parallel trend before the intervention between treatment group and control group, we later employ propensity score matching (see details in Section \ref{sec:time_based_horizontal_analysis}). Additionally, given that tweets treated after the intervention are not the same as those treated before the intervention, we consider confounding factors (\ie, control variables, \eg, content and author characteristics) that may affect the outcome alongside the roll-out of Community Notes feature (see details in Section \ref{sec:confounding_analysis}).} This assumption allows for the estimation of the average treatment effect on the treated group (ATT). Specifically, the model compares the changes in an outcome variable over time for a treatment group to the changes in the same outcome variable over time for a control group. Importantly, the term ``intervention'' in this study refers to the ``roll-out'' of the Community Notes feature, and the term ``treatment'' refers to the display of notes on the corresponding source tweets. 

There are two dummy variables associated to the group level and time period level in the DiD model: (i) The group-specific dummy variable $\mathit{Treated}$ indicates the treatment group ($=$ 1) or control group ($=$ 0). (ii) The time-specific dummy variable $\mathit{Vis}$ indicates the pre-intervention ($=$ 0) or post-intervention period ($=$ 1). Comparing the differences in engagement between the two groups before and after the intervention, the DiD model can help identify the treatment effect of the intervention by estimating the interaction term $\mathit{Treated} \times \mathit{Vis}$. Note that the treatment group includes tweets that receive one or more CRH notes, regardless of the delay in note display ($\mathit{Delay}$). The control group includes the tweets that did not receive any CRH notes. Then, we can use DiD to estimate the extra reduction of average engagement in the treatment group after the intervention, compared to the baseline before the intervention and relative to the control group.

Formally, our DiD model is specified as:
\begin{equation}
\begin{aligned}
log(E(Outcome_{it}|\bm{x^*_{it}})) = \beta_{0} + \beta_{1}Treated_{i} + \beta_{2}Vis_{t} + \beta_{3}Treated_{i} \times Vis_{t} + \bm{\alpha^{'}x_{it}},
\end{aligned}
\end{equation}
where $\mathit{Outcome_{it}}$ indicates the retweet count ($\mathit{RetweetCount_{it}}$) or like count ($\mathit{LikeCount_{it}}$) of tweet $\mathit{i}$ posted at time $\mathit{t}$. Because retweet count and like count are all non-negative count variables with overdispersion, we employ a negative binomial regression. $\mathit{Treated_{i}}$ is the group-specific dummy variable indicating whether the tweet $\mathit{i}$ belongs to the treatment group ($=$ 1) or control group ($=$ 0). $\mathit{Vis_{t}}$ is the time-specific dummy variable indicating whether the tweet $\mathit{i}$ is posted after the roll-out of Community Notes ($=$ 1) or not ($=$ 0). The interaction term, $\mathit{Treated_{i} \times Vis_{t}}$, is difference-in-differences estimation and denotes the treatment effect (\ie, the effect attributed to the roll-out of Community Notes). $\mathit{\bm{x_{it}}}$ is a vector including control variables constructed in Section \ref{sec:variable_extraction}. $\mathit{\bm{x^*_{it}}}$ indicates all the independent variables. $\mathit{\beta_{0}}$ is the intercept. For the sake of interpretability, we $z$-standardize all continuous variables in $\mathit{\bm{x^*_{it}}}$.

\subsubsection{Regression Discontinuity Design (RDD) model}

Given that the detailed engagement timelines before and after the display of CRH notes are not available in our dataset, we cannot use DiD to estimate the treatment effect of Community Notes program at the note display level (\ie, the treatment effect attributed to the display of notes). However, we can use Regression Discontinuity Design (RDD, ref. \cite{angrist_mostly_2009}) as a complementary method to conduct an analysis of the effectiveness of Community Notes at the note display level in terms of aggregated engagement. RDD is a quasi-experimental research design that aims to estimate the effect of a treatment by exploiting a discontinuity in the relationship between a running variable and an outcome variable at a specific threshold or cutoff point. The key rationale behind this is that the expected outcomes from samples with close values on a running variable should exhibit smooth continuity around the cutoff point without the treatment. Therefore, in an RDD model, samples are assigned to the treatment or control group based on their values on a running variable, and the treatment effect is estimated by comparing the outcomes of samples just above and just below the cutoff point \cite{horta_ribeiro_deplatforming_2023,liu_empirical_2022,freire_understanding_2022}. In our analysis, $\mathit{NoteScore}$ is the running variable that determines whether the notes are displayed (\ie, CRH notes or non-CRH notes). We use the values of $\mathit{NoteScore}$ to select tweets just above (treatment group, $\mathit{Treated} = 1$) and just below (control group, $\mathit{Treated} = 0$) the threshold in a narrow window. By estimating the group-specific dummy variable $\mathit{Treated}$, the RDD model can identify the discontinuity jump (treatment effect) from control group to treatment group.

Formally, our RDD model is specified as:
\begin{equation}
\begin{aligned}
log(E(Outcome_{i}|\bm{x^*_{i}})) = \beta_{0} + \beta_{1}Treated_{i} + \beta_{2}Delay_{i} + \beta_{3}NoteScore_{i} + \bm{\alpha^{'}x_{i}},
\end{aligned}
\end{equation}
where $\mathit{Outcome_{i}}$ indicates retweet count or like count of tweet $\mathit{i}$. Different from the DiD model, the RDD model does not take into account the longitudinal time periods before and after the intervention, but rather focuses on the running variable $\mathit{NoteScore_{i}}$ that determines the display of the note. The cutoff point for $\mathit{NoteScore_{i}}$ is set at 0.40, indicating that only notes with a helpfulness score of 0.40 or above are eligible to be displayed on the corresponding tweets. Note that in the RDD model, we only consider tweets with notes decided by the Core model (v1.0) to keep the same standard on the running variable. The selection of the window around the cut-off point is specified in the analysis (see Section \ref{sec:score_based_vertical_analysis}). $\mathit{Treated_{i}}$ captures the discontinuity (\ie, treatment effect) at the point of cutoff. $\mathit{Delay_{i}}$ is the time duration (days) between the posting time of tweet $\mathit{i}$ and the point at which its associated note is displayed to the users. $\bm{\mathit{x_{i}}}$ is a vector including control variables constructed in Section \ref{sec:variable_extraction}. $\bm{\mathit{x^*_{i}}}$ indicates all the independent variables. $\mathit{\beta_{0}}$ is the intercept. We again $z$-standardize all continuous variables in $\bm{\mathit{x_{i}}}$ when fitting the RDD model.

Note that in both models, \ie, DiD and RDD, the treatments are all indicated by the note helpfulness scores (\ie, $\mathit{NoteScore}$) that decide upon the note statuses. However, the two methods incorporate $\mathit{NoteScore}$ and $\mathit{Delay}$ slightly differently. Specifically, at the roll-out level, the DiD model does not incorporate the continuous variable $\mathit{NoteScore}$ as it instead uses a binary treatment indicator (\ie, $\mathit{Treated}$). This binary indicator categorizes tweets into treated and control groups based on the presence or absence of helpful notes, providing a clear distinction for the evaluation of engagement levels before and after the intervention. In contrast, at the note display level, RDD requires specific values of $\mathit{NoteScore}$ to define the cutoff point, set the selection window, and construct treatment and control groups. A similar rationale applies to the variable $\mathit{Delay}$. At the roll-out level, DiD generally evaluates the average engagement level before and after the intervention, comparing the treatment and control groups. The roll-out time of Community Notes feature is thus constant across each treated tweet, irrespective of the delay in note display. The RDD model, on the other hand, focuses on the note display level. The timing of the treatment thus varies among the treated tweets. Consequently, $\mathit{Delay}$ needs to be explicitly included in the RDD model.

\section{Empirical Analysis}

We begin our empirical analysis by summarizing the fundamental changes in notes and fact-checked tweets upon the expansion of Community Notes to the \US and the rest of the world (Section \ref{sec:preliminary_analysis}). We then create treatment and control groups covering time periods before and after the expansions, followed by applying the DiD model to study the effectiveness of Community Notes at the feature roll-out level (Section \ref{sec:time_based_horizontal_analysis}). Subsequently, we present a placebo analysis with different combinations of treatment and control groups (Section \ref{sec:placebo}). Afterwards, we construct treatment and control groups based on the running variable of note helpfulness, and employ the RDD model to estimate the treatment effect at the note display level (Section \ref{sec:score_based_vertical_analysis}). Finally, we consider potential confounding factors across multiple levels to further consolidate the findings of this study (Section \ref{sec:confounding_analysis}).

\subsection{Fact-Checking via Community Notes}
\label{sec:preliminary_analysis}
\begin{figure}
    \centering
    \includegraphics[width=0.5\columnwidth]{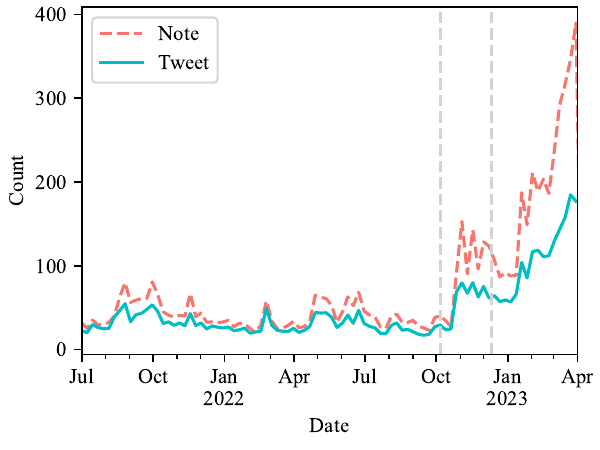}
    \caption{Daily counts of created notes and fact-checked tweets, smoothed by a 7-day moving average. There are two grey vertical dash lines. The first (left) is on October 6, 2022 (Community Notes feature visible in the \US), and the second (right) is on December 11, 2022 (Community Notes feature visible to the world).}
    \label{fig:note_tweet_timeline}
\end{figure}

With X/Twitter expanding Community Notes to the \US and the world, we first perform an initial analysis on the development of the Community Notes feature in terms of contributors, fact-checking notes, and tweets. Of the 64,359 contributors in total, 48,927 contributors (76\%) joined after the global expansion of Community Notes, marking a more than 24-fold increase in the number of contributors during the ``Visible World'' phase (visible to the world) compared to the ``Visible US'' phase (visible only in the \US). Simultaneously, we observe a substantial surge in the daily counts of notes and fact-checked tweets, especially after the expansion to the world (Fig. \ref{fig:note_tweet_timeline}). In addition, we find that the number of notes grows faster than the number of fact-checked tweets. The ratio of notes to tweets increased from 1.37 during the Invisible phase to 1.62 in the ``Visible US'' phase, and further to 1.79 in the ``Visible World'' phase. On average, each tweet received 1.58 notes, with 32.2\% of tweets being fact-checked by more than one note.

We now consider the relationships between the volume of notes and various tweet characteristics (see variables in Section \ref{sec:variable_extraction}). As the majority of tweets receive only a single note, we employ a logistic regression model to analyze the characteristics of tweets that are more likely to receive multiple notes (coded as 1, otherwise 0). Moreover, we model the rating number (\ie, the number of ratings a note receives) with a negative binomial regression. We find that tweets from verified high-influence users with many followers tend to get more notes and ratings (see the coefficients of $\mathit{Verified}$ and $\mathit{Followers}$ in Columns (1) and (2) of Table \ref{tab:summary_statistics}). Tweets with media elements, such as pictures and videos, also have a greater probability of receiving notes and ratings (see the coefficients of $\mathit{Media}$ in Columns (1) and (2) of Table \ref{tab:summary_statistics}).

\begin{figure}
    \centering
    \subfigure[From tweet creation to note creation]{
    \begin{minipage}{0.45\linewidth}
        \centering
        \includegraphics[width=\columnwidth]{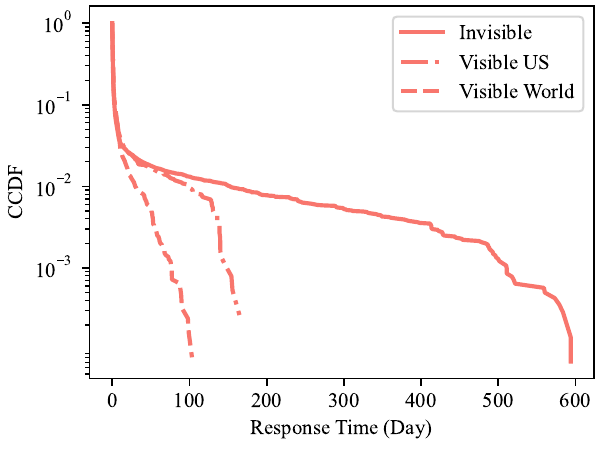}
    \end{minipage}
    }
    \subfigure[From tweet creation to note display]{
    \begin{minipage}{0.45\linewidth}
        \centering
        \includegraphics[width=\columnwidth]{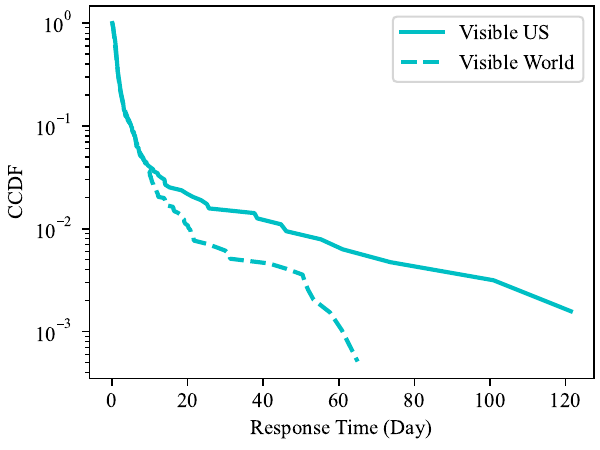}
    \end{minipage}
    }
    \caption{Complementary cumulative distribution functions (CCDFs) of the delays. (a) Delay (response time) between the time of tweet creation and the time of note creation. (b) Delay (response time) between the time of tweet creation and the time when the note is displayed to users.}
    \label{fig:summary_statistics}
\end{figure}

Additionally, we analyze the response time (delay) that notes take to be created and displayed to the users. As shown in Fig. \ref{fig:summary_statistics}(a), the delay from tweets to notes creation has decreased significantly from 5.21 days (before the roll-outs) to 3.08 days (Visible US, $t=3.511, p<0.001$), and to 1.82 days (Visible World, $t=7.763, p<0.001$), on average. When moving the focus to the response time that notes take to be displayed, we find that the delay also decreased (Fig. \ref{fig:summary_statistics}(b)); however, the magnitude of the decrease is relatively small ranging from 2.85 days (Visible US) to 2.23 days (Visible World, $t=2.286, p<0.05$). Notably, these response times are significantly longer than the typical half-life of tweets.\footnote{Previous research found that the median half-life of a tweet, \ie, the time it takes before half of all impressions are created, is about 79.5 minutes \cite{pfeffer_half-life_2023}.} This indicates that the Community Notes program's response time may not be fast enough to display notes during the early stages of diffusion, potentially limiting its effectiveness in significantly reducing engagement with misinformation on X/Twitter.

We also model the response time with linear regression models and find that tweets from verified accounts tend to receive notes faster and with a shorter delay between the creation date and note display (see the coefficients for $\mathit{Verified}$ in Columns (3) and (4) of Table \ref{tab:summary_statistics}).

\begin{table}
\caption{Regression results for response number and time}
\label{tab:summary_statistics}
\begin{tabular}{l*{4}{S}}
    \toprule
    &\multicolumn{2}{c}{Response Number}&\multicolumn{2}{c}{Response Time}\\
    \cmidrule(r){2-3} \cmidrule(r){4-5}
    &{\makecell{(1)\\Note}}&{\makecell{(2)\\Rating}}&{\makecell{(3)\\Creation}}&{\makecell{(4)\\Visible}}\\
    \midrule
    $\mathit{WordNum}$&0.042^{**}&0.020&0.485^{**}&0.215\\
    &(0.015)&(0.020)&(0.179)&(0.138)\\
    $\mathit{Media}$&0.055^{*}&0.155^{***}&0.317&0.155\\
    &(0.028)&(0.034)&(0.329)&(0.293)\\
    $\mathit{Hashtag}$&-0.268^{***}&-0.444^{***}&0.196&-0.031\\
    &(0.048)&(0.052)&(0.570)&(0.294)\\
    $\mathit{Mention}$&-0.614^{***}&-0.680^{***}&0.323&-0.084\\
    &(0.036)&(0.054)&(0.401)&(0.336)\\
    $\mathit{AcctAge}$&-0.029^{*}&0.026&-0.224&-0.157\\
    &(0.014)&(0.016)&(0.137)&(0.121)\\
    $\mathit{Verified}$&0.781^{***}&0.756^{***}&-3.219^{***}&-1.218^{***}\\
    &(0.032)&(0.038)&(0.377)&(0.283)\\
    $\mathit{Followers}$&0.261^{***}&0.277^{***}&0.084&0.047\\
    &(0.016)&(0.013)&(0.173)&(0.169)\\
    $\mathit{Followees}$&0.025^{*}&-0.022^{**}&0.015&-0.012\\
    &(0.012)&(0.008)&(0.059)&(0.051)\\
    Emotions&$\checkmark$&$\checkmark$&$\checkmark$&$\checkmark$\\
    Topics&$\checkmark$&$\checkmark$&$\checkmark$&$\checkmark$\\
    Periods&$\checkmark$&$\checkmark$&$\checkmark$&$\checkmark$\\
    $\mathit{Intercept}$&-1.832^{***}&2.353^{***}&8.042^{***}&3.749^{***}\\
    &(0.039)&(0.042)&(0.564)&(0.472)\\
    Observations&{30,267}&{30,267}&{30,267}&{2,586}\\
    \bottomrule
    \multicolumn{5}{l}{\makecell[l]{Regression models with robust standard errors in parentheses. \\ $^*p<0.05$, $^{**}p<0.01$, and $^{***}p<0.001$.}}\\
\end{tabular}
\end{table}

In sum, our initial analysis reveals a rapid adoption of Community Notes, marked by a significant increase in contributors, notes, and fact-checked tweets, especially after the global expansion. Verified users with many followers and tweets with media elements tend to receive more notes and ratings. Despite a decrease after the roll-outs, the response time from tweet creation to note display is still significantly longer than the typical half-life of tweets, potentially limiting the effectiveness of Community Notes in reducing overall engagement with misinformation on X/Twitter.

\subsection{Analysis at the Feature Roll-Out Level (DiD)}
\label{sec:time_based_horizontal_analysis}

In this section, we employ Difference-in-Differences (DiD) to analyze whether the introduction of the Community Notes feature and its roll-out to users in the \US and around the world has reduced engagement with misinformation on X/Twitter (\ie, at the feature roll-out level). For this purpose, we categorize tweets into three groups based on their note statuses and classifications. The first group, F-CRH, consists of tweets with notes that indicate they are potentially misleading and are currently rated helpful. Notes with CRH status are in the highest priority and displayed on the corresponding tweets directly. Misleading tweets (\ie, misinformation, false information) in the F-CRH group contain at least one CRH note and can also contain other NMR or CRNH notes. The number of tweets in the F-CRH group is 3,387. As of April 13, 2023, Community Notes has different status control rules for notes indicating misleading and not misleading. Notes marking corresponding tweets as ``not misleading'' have no CRH status, even if they have enough ratings and high helpfulness scores. Considering this subset of tweets, we have the second group, T-NMR, which includes tweets with NMR notes indicating that they are not misleading (\ie, true information) and have at least 5 ratings. Excluding tweets in the F-CRH group, the T-NMR group has 5,602 tweets with a high probability of being true. The third group, F-NCRH, includes tweets with notes indicating that they are potentially misleading but do not have a CRH status. Excluding tweets in the F-CRH and T-NMR groups, the F-NCRH group contains 20,669 unique tweets.

Community Notes implemented two interventions to change note visibility settings during the program. On October 6, 2022, the first intervention made notes visible to all users in the \US \cite{twitter_helpful_2022}. Two months after the first intervention, Community Notes expanded note visibility to the whole world on December 11, 2022, marking the second intervention. As of the date of data collection, the second intervention had been in effect for almost four months. Specifically, the two intervention periods are $\mathit{VisUS}$ (06-10-2022 to 10-12-2022) and $\mathit{VisWorld}$ (11-12-2022 to 06-04-2023). To perform parallel tests and longitudinal comparisons between the treatment and control groups, we divide the time before the interventions into five three-month periods, namely, $\mathit{Pre5}$ (01-07-2021 to 30-09-2021), $\mathit{Pre4}$ (01-10-2021 to 30-12-2021), $\mathit{Pre3}$ (01-01-2022 to 31-03-2022), $\mathit{Pre2}$ (01-04-2022 to 30-06-2022), and $\mathit{Pre1}$ (01-07-2022 to 05-10-2022). Notably, we treat the roll-out of the Community Notes to the \US and the rest of the world as two interventions. The reason is that there were significant changes in contributors, fact-checking notes, and tweets along with the global availability of Community Notes (see Section \ref{sec:preliminary_analysis}). Furthermore, even though there was no distinction in visibility to \US users between the two interventions, the Community Notes feature was still evolving and gaining traction among \US users. The impact of Community Notes during different phases may thus exhibit a different effect -- even for \US users. Based on these reasons, we consider the two interventions separately to explore whether the effect of Community Notes after the second intervention is larger than that after the first intervention and to give more detailed estimations.\footnote{As an additional check, we also repeated our analysis without distinguishing between the two expansions of the Community Notes feature (\ie, with a single treatment framework). The results are consistent with our two-treatment examination framework. The additional results are reported in Table \ref{tab:mis_all_did_rollout} in the Appendix.}

Longitudinally, we take F-CRH as the treatment group and F-NCRH as the control group. CRH notes on tweets within the F-CRH group are visible to the \US and the whole world respectively during the two intervention periods (as illustrated in Fig. \ref{fig:did_rdd_groups}(a) in the Appendix). First, we conduct propensity score matching (PSM) to match F-CRH and F-NCRH groups according to the pre-defined variables in Section \ref{sec:variable_extraction}. With the caliper of 0.001, ``common'' support, and ``noreplacement'' settings, we only discard 38 out of 24,056 tweets (0.15\%) to achieve all biases lower than 0.05 across the variables derived from tweets and authors. Second, we test the parallel trends between F-CRH and F-NCRH groups. Before the interventions, the changes in retweet count and like count within F-CRH and F-NCRH groups present a parallel trend and have no statistically significant difference (Fig. \ref{fig:parallel_test}(a) in the Appendix). Ultimately, our investigation proceeds to analyze the impact of the interventions on the dissemination of false or misleading information. Regarding the first intervention, where Community Notes was expanded to the \US, we consider the pre-intervention period as the three-month period leading up to October 6, 2023 ($\mathit{Pre1}$); while the two-month period from October 6, 2023 to December 11, 2022 is taken as the post-intervention period ($\mathit{VisUS}$). The coefficients of $\mathit{FCRH \times Vis}$ are all significantly negative in Columns (1) and (3) of Table \ref{tab:mis_all_did} ($\mathit{coef.}=-0.439, p<0.01$ in Column (1); $\mathit{coef.}=-0.403, p<0.05$ in Column (3)). This indicates that F-CRH tweets received 35.5\% fewer retweets (95\%~CI: $[-53.1\%, -11.4\%]$) and 33.2\% fewer likes (95\%~CI: $[-51.9\%, -7.2\%]$) compared to the baseline before the first intervention and relative to the F-NCRH tweets. Community Notes thus seems to reduce the engagement with misinformation after the CRH notes are visible to the \US from the perspectives of retweet count and like count.

\begin{figure*}
    \centering
    \subfigure[Two-week rolling averages of retweet count]{
    \begin{minipage}{0.45\linewidth}
        \centering
        \includegraphics[width=\columnwidth]{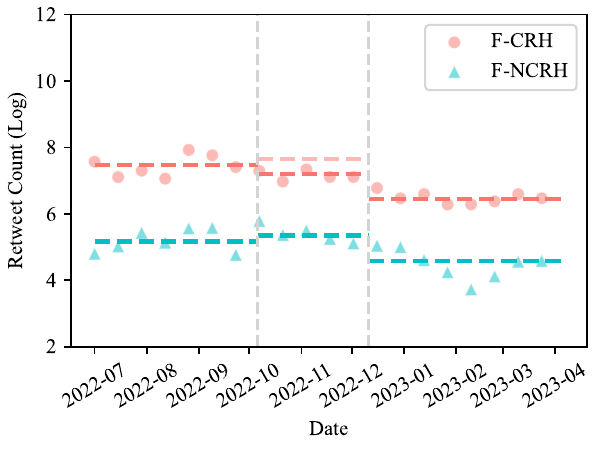}
    \end{minipage}
    }
    \subfigure[Two-week rolling averages of like count]{
    \begin{minipage}{0.45\linewidth}
        \centering
        \includegraphics[width=\columnwidth]{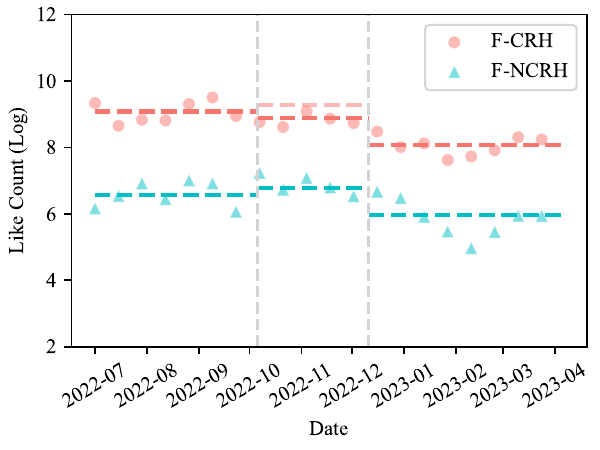}
    \end{minipage}
    }
    \caption{Two-week rolling averages of retweet count and like count in treatment (F-CRH) and control (F-NCRH) groups from 2022-07 to 2023-04. The two vertical grey dash lines in each figure indicate two dates, October 6, 2022 (left) and December 11, 2022 (right), when Community Notes was expanded to the \US and the world respectively. (a) The points are 2-week rolling averages of retweet count. The horizontal dash lines represent the averages of retweet count during different time periods, and illustrate the results of DiD models in Columns (1) and (2) of Table \ref{tab:mis_all_did}. The lighter red dash line during $\mathit{VisUS}$ denotes the estimated average without the treatment effect ($\mathit{FCRH \times Vis} = -0.439, p < 0.01$) in Column (1) of Table \ref{tab:mis_all_did}. (b) The points are 2-week rolling averages of like count. The horizontal dash lines represent the averages of like count during different time periods, and illustrate the results of DiD models in Columns (3) and (4) of Table \ref{tab:mis_all_did}. The lighter red dash line during $\mathit{VisUS}$ denotes the estimated average without the treatment effect ($\mathit{FCRH \times Vis} = -0.403, p < 0.05$) in Column (3) of Table \ref{tab:mis_all_did}.}
    \label{fig:mis_did}
\end{figure*}

Continuing our investigation, we further scrutinize the effect of the second intervention, \ie, the global expansion of the Community Notes feature. Here, the pre-intervention phase is defined as the two-month period ranging from October 6, 2023 to December 11, 2022 ($\mathit{VisUS}$). The post-intervention phase, on the other hand, is characterized by the duration of four months starting on December 11, 2022 and ending on April 6, 2023 ($\mathit{VisWorld}$). As shown in Columns (2) and (4) of Table \ref{tab:mis_all_did}, in the second intervention, the coefficients of $\mathit{FCRH \times Vis}$ for both retweet count and like count are found to be statistically non-significant. This suggests that the impact of Community Notes feature on reducing engagement with false or misleading information after the second intervention does not exhibit a significant difference compared to that observed after the first intervention. Fig \ref{fig:mis_did} displays the variations in the retweet and like counts within the F-CRH and F-NCRH groups across both the pre- and post-intervention periods. During the first intervention, there are slight decreases in both retweet count and like count within the F-CRH group. In the absence of the first intervention, it is anticipated that the engagement with F-CRH tweets should go like the lighter red dash line (according to the coefficients of $\mathit{Vis}$ in Columns (1) and (3) of Table \ref{tab:mis_all_did}). The differences between the two red dash lines in Figs. \ref{fig:mis_did}(a) and (b) during the first prevention period can be regarded as the treatment effects. In the second intervention, there is a significant decrease in the engagement with tweets throughout both F-CRH and F-NCRH groups (according to the coefficients of $\mathit{Vis}$ in Columns (2) and (4) of Table \ref{tab:mis_all_did}). However, the non-significant coefficients of $\mathit{FCRH \times Vis}$ for both retweet count and like count indicate that no special treatment effect on the F-CRH group is captured (Columns (2) and (4) of Table \ref{tab:mis_all_did}).

In summary, our findings so far indicate that the first intervention, involving the expansion of Community Notes to the \US, is linked to decreases in engagement with misinformation, as evidenced by reduced retweet and like counts. The second intervention, \ie, the global roll-out, does not yield a significant increase in the treatment effect as compared to the first intervention.

\begin{table}
\caption{DiD regression results for retweet count and like count within F-CRH (treatment) and F-NCRH (control) groups.}
\label{tab:mis_all_did}
\begin{tabular}{l*{4}{S}}
\toprule
&\multicolumn{2}{c}{Retweet Count}&\multicolumn{2}{c}{Like Count}\\
\cmidrule(r){2-3} \cmidrule(r){4-5}
&{\makecell{(1)\\Pre-US}}&{\makecell{(2)\\US-World}}&{\makecell{(3)\\Pre-US}}&{\makecell{(4)\\US-World}}\\
\midrule
		$\mathit{FCRH}$&1.172^{***}&0.855^{***}&1.212^{***}&0.944^{***}\\
		&(0.132)&(0.120)&(0.139)&(0.115)\\
		$\mathit{Vis}$&0.177^{*}&-0.761^{***}&0.202*&-0.809^{***}\\
		&(0.080)&(0.065)&(0.085)&(0.072)\\
		$\mathit{FCRH \times Vis}$&-0.439^{**}&-0.123&-0.403^{*}&-0.127\\
		&(0.162)&(0.133)&(0.168)&(0.132)\\
		$\mathit{WordNum}$&0.081^{*}&0.134^{***}&-0.071&-0.020\\
		&(0.038)&(0.031)&(0.042)&(0.033)\\
		$\mathit{Media}$&0.515^{***}&0.687^{***}&0.453^{***}&0.509^{***}\\
		&(0.078)&(0.057)&(0.084)&(0.063)\\
		$\mathit{Hashtag}$&-0.872^{***}&-0.722^{***}&-1.070^{***}&-0.970^{***}\\
		&(0.150)&(0.096)&(0.141)&(0.096)\\
		$\mathit{Mention}$&-1.188^{***}&-1.291^{***}&-1.163^{***}&-1.302^{***}\\
		&(0.102)&(0.072)&(0.110)&(0.075)\\
		$\mathit{AcctAge}$&-0.021&-0.018&-0.002&0.012\\
		&(0.034)&(0.025)&(0.033)&(0.028)\\
		$\mathit{Verified}$&0.596^{***}&0.821^{***}&0.502^{***}&0.783^{***}\\
		&(0.083)&(0.061)&(0.083)&(0.065)\\
		$\mathit{Followers}$&0.159^{***}&0.179^{***}&0.199^{***}&0.221^{***}\\
		&(0.015)&(0.013)&(0.014)&(0.012)\\
		$\mathit{Followees}$&0.037&0.028&0.041&0.030*\\
		&(0.036)&(0.016)&(0.039)&(0.012)\\
		Emotions&$\checkmark$&$\checkmark$&$\checkmark$&$\checkmark$\\
		Topics&$\checkmark$&$\checkmark$&$\checkmark$&$\checkmark$\\
		$\mathit{Intercept}$&7.062^{***}&6.899^{***}&8.820^{***}&8.658^{***}\\
		&(0.099)&(0.080)&(0.109)&(0.090)\\
		Observations&{5,019}&{11,821}&{5,019}&{11,821}\\
\bottomrule
\multicolumn{5}{l}{\makecell[l]{Negative binomial regression models with robust standard errors \\ in parentheses. $^*p<0.05$, $^{**}p<0.01$, and $^{***}p<0.001$.}}\\
\end{tabular}
\end{table}
 
\subsection{Placebo Analysis}
\label{sec:placebo}

We execute two placebo tests to further explore the effect of Community Notes in reducing engagement with misinformation. The special group, T-NMR, is taken into the placebo tests as a mediating group. In the first test, we replace F-CRH with T-NMR as the virtual treatment group and compared it to F-NCRH as the control group. In the second test, we use T-NMR as the control group and compare it to F-CRH as the treatment group.

\subsubsection{T-NMR VS F-NCRH} 
First, we conduct propensity score matching again with the same parameters as detailed in Section \ref{sec:time_based_horizontal_analysis} to get matched T-NMR and F-NCRH groups with discarding 255 out of 26,271 tweets (0.97\%). The biases of tweet- and author-related variables between the matched groups are all lower than 0.05. Second, we perform a parallel test between the treatment and control groups during the periods before and after the interventions. Prior to the interventions, both groups demonstrate parallel trends in misinformation diffusion in terms of retweet count and like count (Fig. \ref{fig:parallel_test}(b) in the Appendix). Moreover, the unparalleled changes in T-NMR and F-NCRH groups during the two periods following the interventions are even more significant than those in F-CRH and F-NCRH groups (Fig. \ref{fig:parallel_test}(a) in the Appendix). We then examine the virtual treatment effects using the DiD model. The coefficients of $\mathit{TNMR \times Vis}$ in Columns (1) and (3) of Table \ref{tab:placebo_test_did} in the Appendix are all statistically significantly negative ($\mathit{coef.}=-0.503, p<0.05$ in Column (1); $\mathit{coef.}=-0.439, p<0.05$ in Column (3)), just like those in Columns (1) and (3) of Table \ref{tab:mis_all_did}. Specifically, T-NMR tweets received 39.5\% fewer retweets (95\%~CI: $[-59.5\%, -9.9\%]$) and 35.5\% fewer likes (95\%~CI: $[-55.5\%, -6.6\%]$) compared to the baseline before the first intervention and relative to the F-NCRH tweets. This indicates that with the expansion of Community Notes to the \US, engagement with tweets in the T-NMR group also decrease even though they have no CRH notes. Also, the coefficients of $\mathit{TNMR \times Vis}$ are both not significant during the second intervention (Columns (2) and (4) of Table \ref{tab:placebo_test_did} in the Appendix). With the F-NCRH group as the control group to compare, F-CRH and T-NMR groups show the same results regarding the effects of the interventions. This suggests that the decrease in the diffusion of misinformation may not be solely attributed to the targeted treatment of Community Notes on misleading tweets. Instead, it may be influenced by broader changes associated with the expansion of Community Notes, which could have impacted both misleading and non-misleading tweets in a similar manner.

\subsubsection{F-CRH VS T-NMR}
\begin{figure}
    \centering
    \subfigure[Two-week rolling averages of retweet count]{
    \begin{minipage}{0.45\linewidth}
        \centering
        \includegraphics[width=\columnwidth]{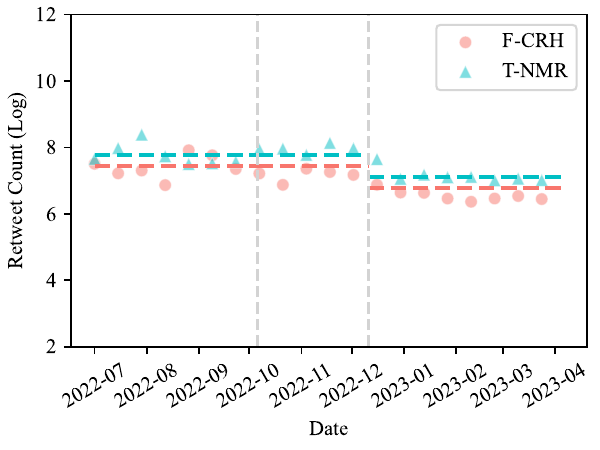}
    \end{minipage}
    }
    \subfigure[Two-week rolling averages of like count]{
    \begin{minipage}{0.45\linewidth}
        \centering
        \includegraphics[width=\columnwidth]{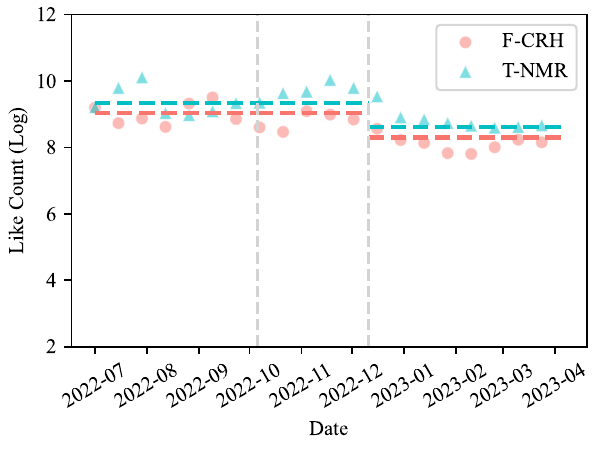}
    \end{minipage}
    }
    \caption{Two-week rolling averages of retweet count and like count in treatment (F-CRH) and control (T-NMR) groups from 2022-07 to 2023-04. The two vertical grey dash lines in each figure indicate two dates, October 6, 2022 (left) and December 11, 2022 (right), when Community Notes was expanded to the \US and the world respectively. (a) The points are 2-week rolling averages of retweet count. The horizontal dash lines represent the averages of retweet count during different time periods, and illustrate the results of DiD models in Columns (1) and (2) of Table \ref{tab:tf_high_rank_did}. The two lines are always parallel over time, which corresponds to the statistically non-significant coefficient estimates of $\mathit{FCRH \times Vis}$ in Columns (1) and (2). (b) The points are 2-week rolling averages of like count. The horizontal dash lines represent the averages of like count during different time periods, and illustrate the results of DiD models in Columns (3) and (4) of Table \ref{tab:tf_high_rank_did}. The two lines are always parallel over time, which corresponds to the statistically non-significant coefficient estimates of $\mathit{FCRH \times Vis}$ in Columns (3) and (4).}
    \label{fig:tf_did}
\end{figure}

When comparing to the F-NCRH group, the F-CRH (misleading) and T-NMR (not misleading) groups show similar results. We then take T-NMR as the control group to examine the impact of the interventions on the engagement with misinformation within the F-CRH group. Based on the previous analysis in this study, it is expected that there would be no significant treatment effect on the F-CRH group when comparing the disparities between the F-CRH and T-NMR groups. To test this hypothesis, we first match the treatment and control groups through propensity score matching with the same parameters as described in Section \ref{sec:time_based_horizontal_analysis}. We discard 490 out of 8,989 (5.5\%) tweets to get matched groups with all variable biases lower than 0.05. We then perform a parallel test between the F-CRH and T-NMR groups across time periods. As expected, they have parallel trends regardless of the interventions (Fig. \ref{fig:parallel_test}(c) in the Appendix). Using the DiD model, we further validate that there is no significant treatment effect for the two interventions (the coefficients of $\mathit{FCRH \times Vis}$ in the four columns of Table \ref{tab:tf_high_rank_did}). As observed in Fig. \ref{fig:tf_did}, in addition to the parallel trends during the first intervention, the decreases in retweet count and like count within both F-CRH and T-NMR groups are no longer statistically significant (the coefficients of $\mathit{Vis}$ in Columns (1) and (3) of Table \ref{tab:tf_high_rank_did}). During the second intervention, the engagement with misleading and not misleading tweets significantly decreases in a parallel way, aligning with the previous findings in this study (the coefficients of $\mathit{Vis}$ in Columns (2) and (4) of Table \ref{tab:tf_high_rank_did}).

Additionally, we find that misleading tweets get not only fewer retweets (Fig. \ref{fig:ccdf_tf}(a) in the Appendix, $KS=0.121, p<0.001$), which supports earlier work conducted during the pilot phase of Community Notes \cite{drolsbach_diffusion_2023}, but also fewer likes than not misleading tweets (Fig. \ref{fig:ccdf_tf}(b) in the Appendix, $KS=0.133, p<0.001$). Furthermore, the T-NMR group consistently exhibits slightly higher retweet and like counts compared to the F-CRH group, regardless of the interventions (Fig. \ref{fig:tf_did}). The differences still exist under the control of other covariates (the coefficients of $\mathit{FCRH}$ in Table \ref{tab:tf_high_rank_did}). The distributions of retweet count and like count in misleading and not misleading tweets across different time periods are shown in Fig \ref{fig:ccdf_tf_phases} in the Appendix.

\textbf{Consideration of the fact-checking delay:}
The above results do not provide evidence supporting that the roll-out of Community Notes specifically reduced engagement with misinformation in terms of retweet count and like count. Nevertheless, the effectiveness of Community Notes may be sensitive to fact-checking delays. We observed that the response time to note display is significantly longer than the impression half-life of source tweets (see Section \ref{sec:preliminary_analysis}), which suggests that notes were displayed after the most viral stages of diffusion. This delay has the potential to diminish the effect of Community Notes feature on reducing the aggregated engagement. Given this, we limit our analysis to a subgroup of tweets that received CRH notes in a relatively short delay and reevaluate the results. Notably, the shortest delay for note display in the dataset is 80.2 minutes, which is still longer than the half-life. Therefore, we lack sufficient samples to conduct the subgroup analysis around the half-life. Instead, we consider tweets that received CRH notes within 0.569 day (25\% percentile) in the F-CRH treatment group, and the control group is still T-NMR. As reported in Table \ref{tab:tf_high_rank_did_delay} of the Appendix, the coefficients of $\mathit{FCRH \times Vis}$ in the four columns are all not statistically significant, which aligns with the results obtained without the limitation of the delay in note display.

Altogether, the results from the placebo tests imply that the roll-out of Community Notes did not lead to a statistically significant reduction in engagement with misinformation relative to truthful information. Instead, the diffusion of both true and false fact-checked tweets significantly decreased in parallel after the roll-out, which may be due to broader changes on the platform affecting both misleading and non-misleading tweets in a similar manner. This finding persists even when we restrict our analysis to tweets with a relatively short delay in note display. Furthermore, we obtain qualitatively identical findings when repeating the analysis without distinguishing between the two expansions of the Community Notes feature (see Appendix, Table \ref{tab:placebo_rollout}).

\begin{table}
  \caption{DiD regression results for retweet count and like count within F-CRH (treatment) and T-NMR (control) groups.}
  \label{tab:tf_high_rank_did}
  \begin{tabular}{l*{4}{S}}
    \toprule
    &\multicolumn{2}{c}{Retweet Count}&\multicolumn{2}{c}{Like Count}\\
    \cmidrule(r){2-3} \cmidrule(r){4-5}
    &{\makecell{(1)\\Pre-US}}&{\makecell{(2)\\US-World}}&{\makecell{(3)\\Pre-US}}&{\makecell{(4)\\US-World}}\\
    \midrule
    $\mathit{FCRH}$&-0.289&-0.403^{***}&-0.320^{*}&-0.327^{**}\\
    &(0.155)&(0.104)&(0.149)&(0.106)\\
    $\mathit{Vis}$&-0.177&-0.665^{***}&-0.130&-0.729^{***}\\
    &(0.132)&(0.083)&(0.129)&(0.080)\\
    $\mathit{FCRH \times Vis}$&-0.184&0.029&-0.118&-0.022\\
    &(0.176)&(0.117)&(0.177)&(0.123)\\
    $\mathit{WordNum}$&-0.060&0.047&-0.237^{***}&-0.122^{***}\\
    &(0.045)&(0.030)&(0.047)&(0.034)\\
    $\mathit{Media}$&-0.079&0.286^{***}&-0.078&0.127^{*}\\
    &(0.092)&(0.057)&(0.096)&(0.063)\\
    $\mathit{Hashtag}$&-0.328&-0.488^{***}&-0.530^{**}&-0.675^{***}\\
    &(0.226)&(0.114)&(0.190)&(0.119)\\
    $\mathit{Mention}$&-0.478^{***}&-0.585^{***}&-0.531^{***}&-0.625^{***}\\
    &(0.118)&(0.073)&(0.124)&(0.073)\\
    $\mathit{AcctAge}$&-0.060&-0.087^{**}&-0.018&-0.092^{**}\\
    &(0.041)&(0.028)&(0.042)&(0.030)\\
    $\mathit{Verified}$&-0.087&0.279^{***}&-0.127&0.289^{***}\\
    &(0.111)&(0.063)&(0.111)&(0.066)\\
    $\mathit{Followers}$&0.251^{***}&0.296^{***}&0.285^{***}&0.331^{***}\\
    &(0.030)&(0.022)&(0.029)&(0.019)\\
    $\mathit{Followees}$&-0.002&-0.016&0.042&0.002\\
    &(0.064)&(0.012)&(0.065)&(0.013)\\
    Emotions&$\checkmark$&$\checkmark$&$\checkmark$&$\checkmark$\\
    Topics&$\checkmark$&$\checkmark$&$\checkmark$&$\checkmark$\\
    $\mathit{Intercept}$&9.163^{***}&8.484^{***}&10.919^{***}&10.310^{***}\\
    &(0.164)&(0.099)&(0.156)&(0.098)\\
    Observations&{1,716}&{6,463}&{1,716}&{6,463}\\
    \bottomrule
    \multicolumn{5}{l}{\makecell[l]{Negative binomial regression models with robust standard errors \\ in parentheses. $^*p<0.05$, $^{**}p<0.01$, and $^{***}p<0.001$.}}\\
\end{tabular}
\end{table}

\subsection{Analysis at the Note Display Level (RDD)}
\label{sec:score_based_vertical_analysis}
In the previous sections, we focused on different intervention periods and analyzed whether the introduction of the Community Notes feature and its roll-out to users in the \US and around the world has reduced engagement with misinformation on X/Twitter (\ie, at the feature roll-out level). Next, we use our RDD model as a complementary method to study the effectiveness of the Community Notes feature at the note display level, that is, whether there is a reduction in engagement with misleading posts on which notes are displayed compared to those on which notes are not displayed. 

To estimate our RDD model, we take $\mathit{NoteScore}$ as the running variable to divide tweets into treatment and control groups with a cutoff point of 0.40. When fitting the RDD model, we only consider tweets posted after the roll-out intervention (roll-out period, as illustrated in Fig \ref{fig:did_rdd_groups}(b) in the Appendix). The treatment group, F-CRH, includes tweets with CRH notes that indicate they are potentially misleading and have helpfulness scores of no less than 0.40. The number of tweets in the F-CRH group is 2,133. Notably, if one F-CRH tweet has multiple CRH notes, the $\mathit{NoteScore}$ is the average score of all its CRH notes. The control group, F-NCRH, includes tweets with non-CRH (NCRH) notes that indicate they are potentially misleading and have helpfulness scores of less than 0.40. Excluding tweets in the F-CRH group, F-NCRH has 9,671 unique tweets. Similarly, if one F-NCRH tweet has multiple NCRH notes, the $\mathit{NoteScore}$ is the average score of all its NCRH notes.

\begin{figure*}
    \centering
    \subfigure[Averages of retweet count]{
    \begin{minipage}{0.45\linewidth}
        \centering
        \includegraphics[width=\columnwidth]{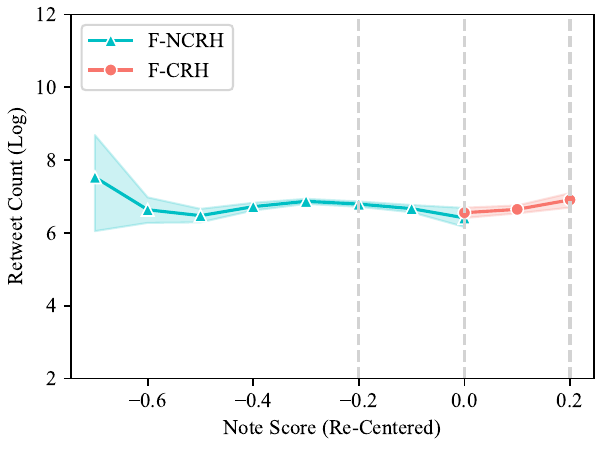}
    \end{minipage}
    }
    \subfigure[Averages of like count]{
    \begin{minipage}{0.45\linewidth}
        \centering
        \includegraphics[width=\columnwidth]{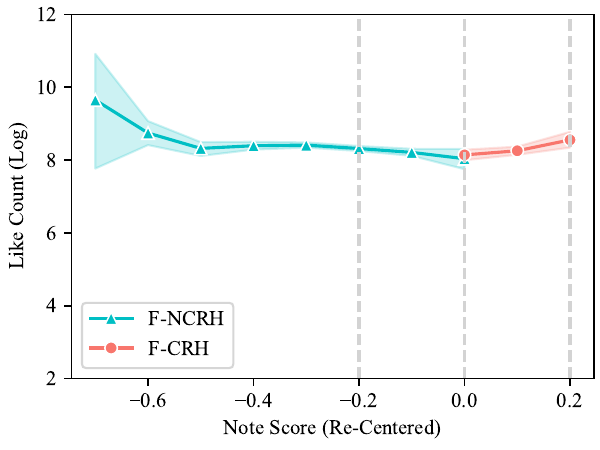}
    \end{minipage}
    }
    \caption{The averages of retweet count and like count across the note score points. Note scores are re-centered based on the cutoff point of 0.40. The shadow areas represent 95\% Confidence Intervals. (a) The averages of retweet count across the note score points. There is no obvious discontinuity on the cutoff point. (b) The averages of like count across the note score points. There is no obvious discontinuity on the cutoff point.}
    \label{fig:mis_rdd}
\end{figure*}

We re-center $\mathit{NoteScore}$, and get the average retweet and like counts on different note score points. As depicted in Fig. \ref{fig:mis_rdd}, there is no clear discontinuity at the cutoff point. To test the discontinuity with the RDD model, we first set the window width as 0.20. The symmetrical distributions of retweet count and like count in the selected window are observed in Figs. \ref{fig:mis_rdd}(a) and (b). Next, we conduct propensity score matching with the same settings as detailed in Section \ref{sec:time_based_horizontal_analysis}. As a result, 325 out of 5,255 tweets (6.2\%) are omitted, and all the variable biases between the two groups are lower than 0.05. Finally, we use the matched groups to perform RDD analysis. The coefficients of $\mathit{Treated}$ in Columns (1) and (3) of Table \ref{tab:vis_rdd} are not statistically significant. This indicates that there is no discontinuity at the cutoff point, which is consistent with the visual continuity at the cutoff point shown in Fig. \ref{fig:mis_rdd}. Moreover, despite limiting our analysis to tweets posted after the second intervention, during which the engagement with all fact-checked tweets shows a notable decrease (Fig \ref{fig:tf_did}), we still do not observe a significant effect of Community Notes on reducing the engagement with misinformation (the coefficients of $\mathit{Treated}$ in Columns (2) and (4) of Table \ref{tab:vis_rdd}). In addition, the coefficients of $\mathit{Delay}$ in Table \ref{tab:vis_rdd} are not significant or even negative (Column (4) of Table \ref{tab:vis_rdd}). This implies that the fact-checking speed did not have a significant effect in reducing the engagement with fact-checked tweets. Notably, this result is also consistent with our findings in the DiD analysis and previous work \cite{drolsbach_diffusion_2023}.

Altogether, the results of our RDD analysis provide no evidence to support the effectiveness of Community Notes in reducing engagement with misinformation in terms of retweet and like counts, which is consistent with our DiD analysis.

\begin{table}
  \caption{RDD regression results for retweet count and like count within F-CRH (treatment) and F-NCRH (control) groups.}
  \label{tab:vis_rdd}
  \begin{tabular}{l*{4}{S}}
    \toprule
    &\multicolumn{2}{c}{Retweet Count}&\multicolumn{2}{c}{Like Count}\\
    \cmidrule(r){2-3} \cmidrule(r){4-5}    
    &{\makecell{(1)\\Roll-Out}}&{\makecell{(2)\\World}}&{\makecell{(3)\\Roll-Out}}&{\makecell{(4)\\World}}\\
    \midrule
    $\mathit{Treated}$&0.157&0.114&0.180&0.044\\
    &(0.121)&(0.134)&(0.132)&(0.147)\\
    $\mathit{Delay}$&0.018&-0.087&-0.020&-0.133^{*}\\
    &(0.033)&(0.053)&(0.035)&(0.053)\\
    $\mathit{NoteScore}$&-0.871&-0.935&-0.303&-0.216\\
    &(0.506)&(0.562)&(0.562)&(0.623)\\
    $\mathit{WordNum}$&0.038&0.076^{*}&-0.112^{**}&-0.070\\
    &(0.031)&(0.035)&(0.035)&(0.040)\\
    $\mathit{Media}$&0.312^{***}&0.423^{***}&0.181^{**}&0.271^{***}\\
    &(0.055)&(0.062)&(0.060)&(0.068)\\
    $\mathit{Hashtag}$&-0.742^{***}&-0.768^{***}&-0.945^{***}&-1.031^{***}\\
    &(0.118)&(0.126)&(0.118)&(0.130)\\
    $\mathit{Mention}$&-0.669^{***}&-0.853^{***}&-0.734^{***}&-0.916^{***}\\
    &(0.091)&(0.088)&(0.096)&(0.091)\\
    $\mathit{AcctAge}$&-0.003&-0.012&0.013&0.001\\
    &(0.028)&(0.030)&(0.031)&(0.035)\\
    $\mathit{Verified}$&0.142^{*}&0.258^{***}&0.087&0.231^{**}\\
    &(0.065)&(0.073)&(0.070)&(0.082)\\
    $\mathit{Followers}$&0.264^{***}&0.282^{***}&0.335^{***}&0.326^{***}\\
    &(0.026)&(0.040)&(0.027)&(0.033)\\
    $\mathit{Followees}$&0.016&0.019&0.040^{*}&0.028\\
    &(0.023)&(0.024)&(0.020)&(0.019)\\
    Emotions&$\checkmark$&$\checkmark$&$\checkmark$&$\checkmark$\\
    Topics&$\checkmark$&$\checkmark$&$\checkmark$&$\checkmark$\\
    $\mathit{Intercept}$&7.629^{***}&7.369^{***}&9.401^{***}&9.163^{***}\\
    &(0.104)&(0.116)&(0.109)&(0.127)\\
    Observations&{4,930}&{3,951}&{4,930}&{3,951}\\
    \bottomrule
    \multicolumn{5}{l}{\makecell[l]{Negative binomial regression models with robust standard errors \\ in parentheses. $^*p<0.05$, $^{**}p<0.01$, and $^{***}p<0.001$.}}\\
\end{tabular}
\end{table}

\subsection{Confounding Analyses}
\label{sec:confounding_analysis}

To consolidate the findings of this study, we conduct additional analyses to incorporate potential confounding factors that are associated with the expansion of Community Notes on X/Twitter from multiple levels including the platform, Community Notes contributors, authors of tweets, and the tweets themselves.

\subsubsection{Platform and Contributors}
From the platform level, Elon Musk concluded the acquisition of X/Twitter on October 27, 2022, which happened during the first expansion of Community Notes. After the acquisition, Musk made some substantial changes on X/Twitter, which might affect the experience of users and the engagement of content on X/Twitter \cite{conger_how_2023}. This study uses a dummy variable, $\mathit{Musk}$, to control the possible effect of the acquisition of Elon Musk ($=$ 1 if the tweet is posted after the acquisition, otherwise $=$ 0). Moreover, we consider Month-Year fixed effects to further control for other potential time-related confounding factors. In addition, at the Community Notes feature level, with its expansion to the public, many new contributors are enrolled and write notes. Specifically, in the user enrollment dataset of this study, we find an enrollment of 48,928 new contributors (76\% of all contributors) in Community Notes after its expansion to the world. Given the possible selection bias between old contributors in the pilot phase and new contributors enrolled after the expansion, we use another dummy variable, $\mathit{OldWriter}$, to control for the potential selection bias from contributors ($=$ 1 if the tweet is noted by the contributors who had written notes during the pilot phase, otherwise $=$ 0). With the two dummy variables, $\mathit{Musk}$ and $\mathit{OldWriter}$, and Month-Year fixed effects included in the regressions, the coefficients and significance levels of $\mathit{FCRH \times Vis}$ are stable in Table \ref{tab:mis_all_did_ext} in the Appendix, compared to Table \ref{tab:mis_all_did}. The coefficients of $\mathit{Vis}$ and all the Month-Year fixed effects in Columns (1) and (3) of Table \ref{tab:mis_all_did_ext} are not statistically significant. This means that the changes in retweet count and like count before and after the expansion of Community Notes to the \US, although slightly significant in Table \ref{tab:mis_all_did}, can be ignored. In Columns (2) and (4) of Table \ref{tab:mis_all_did_ext}, the coefficients of $\mathit{Vis}$ also become statistically non-significant, but the Month-Year fixed effects are all significantly negative during the months when Community Notes is open to the world. The statistically significant negativity of $\mathit{Vis}$ transfers to the fine-grained Month-Year fixed effects in Columns (2) and (4) of Table \ref{tab:mis_all_did_ext} compared to Table \ref{tab:mis_all_did}. Particularly, the negative Month-Year fixed effects in 2023/02 ($\mathit{coef.}=-1.186, p<0.001$ in Column (2); $\mathit{coef.}=-1.400, p<0.001$ in Column (4)) and 2023/03 ($\mathit{coef.}=-1.229, p<0.001$ in Column (2); $\mathit{coef.}=-1.389, p<0.001$ in Column (4)) are stronger than those in 2023/01 ($\mathit{coef.}=-0.785, p<0.01$ in Column (2); $\mathit{coef.}=-0.922, p<0.001$ in Column (4)) and 2023/04 ($\mathit{coef.}=-1.055, p<0.001$ in Column (2); $\mathit{coef.}=-1.355, p<0.001$ in Column (4)). Month-Year fixed effects capture unobserved confounding factors that have significant impacts on the decrease of information diffusion and vary across different months. Additionally, the coefficients of $\mathit{Musk}$ and $\mathit{OldWriter}$ are not significant across all the columns in Table \ref{tab:mis_all_did_ext}. This indicates that the acquisition and possible selection bias between new and old note authors have no substantial effect on the engagement with tweets noted by the Community Notes.

Furthermore, with the expansion of Community Notes, more and more X/Twitter users can transfer from retweeters to contributors who write or rate notes. This shift in user behavior has the potential to decrease the count of retweets or likes on individual tweets. To address the possible user transition, we redesign the dependent variables by incorporating the note count and rating count for each tweet. Specifically, the new retweet metric, $\mathit{RetweetCountPlus}$, denotes the combined count of retweets, notes, and ratings. Similarly, the new like metric, $\mathit{LikeCountPlus}$, represents the combined count of likes, notes, and ratings. In addition, we incorporate quote tweets (retweets with comments) into the retweets category and reexamine the regressions using the new metric $\mathit{RetweetQuoteCount}$. Using these new dependent variables, we regress F-CRH and F-NCRH groups again with the same independent variables. We find that the Month-Year fixed effects are still statistically significantly negative from 2023/01 to 2023/04 (see Appendix, Table \ref{tab:mis_all_did_ext_plus}). Notably, for both retweet count and like count, the Month-Year effects from 2022/08 to 2022/12 are all statistically non-significant (see Appendix, Table \ref{tab:mis_all_did_ext} and Table \ref{tab:mis_all_did_ext_plus}). This suggests that the engagement with tweets in Community Notes has experienced no statistically significant change associated with unobserved time-specific factors before the second intervention. This finding is concordant with the horizontal trend of retweet count and like count from 2022/07 to 2022/12 in Fig. \ref{fig:mis_did}. The results for other key variables, such as $\mathit{FCRH}$ and $\mathit{FCRH \times Vis}$, are also robust.

\subsubsection{Authors and Tweets}
In this study, we use PSM to balance the samples between treatment and control groups based on the characteristics from the author profile and tweet content. However, it is important to note that the differences in the samples across different time periods may be not fully balanced. We further consider the potential influence of unbalanced factors during different time intervals. Based on matched F-CRH and F-NCRH groups, we further match tweets posted before (during $\mathit{Pre1}$) and after (during $\mathit{VisUS}$) the expansion of Community Notes feature to the \US on October 6, 2022 \cite{twitter_helpful_2022}. 1,214 out of 5,028 (24.1\%) tweets are omitted, and the mean bias of the variables is 0.017. Similarly, we match tweets posted before (during $\mathit{VisUS}$) and after (during $\mathit{VisWorld}$) the expansion of Community Notes feature to the whole world. 5,836 out of 11,842 (49.3\%) tweets are omitted, and the mean bias of the variables is 0.014. We do the same regression analysis as presented in Table \ref{tab:mis_all_did_ext} in the Appendix using the newly matched tweets, and the results remain consistent and stable (see Appendix, Table \ref{tab:mis_all_did_ext_matched}).

Although we mitigate the differences in author profiles among different groups in terms of account age, verified status, follower count, and following count, there are still some unobserved factors that are associated with authors and can affect engagement with online information. To control the unobserved factors related to the authors, we only consider the tweets posted by the same users over different time periods. Additionally, we also need to consider the change in the engagement with content at the platform level to analyze whether the engagement with tweets with notes (included in the Community Notes) is different from those without notes (out of Community Notes).

To do so, we use F-CRH as the treatment group, which includes 3,387 tweets posted from 2,153 unique authors. First, we collect 374,601 new tweets posted by the same authors in the F-CRH group during the three time periods, namely, $\mathit{Pre1}$, $\mathit{VisUS}$, and $\mathit{VisWorld}$ (from 01-07-2022 to 06-04-2023). The new tweets without notes represent the group OCN (Out of Community Notes) and undergo the same preprocessing steps as described in Section \ref{sec:variable_extraction}. Second, during $\mathit{Pre1}$, 210 authors are included in the F-CRH group. We use tweets posted by these authors in F-CRH and OCN groups to analyze the first intervention between $\mathit{Pre1}$ and $\mathit{VisUS}$. Specifically, there are 313 tweets in F-CRH and 11,427 tweets in OCN. Given the substantial disparity in tweet counts between OCN and F-CRH groups, we conduct one-to-one matching with a narrower caliper of 0.0001 to ensure precise matches. As a result, 269 out of 313 tweets (85.9\%) in the F-CRH group have corresponding matches in the OCN group. Third, during $\mathit{VisUS}$, 515 authors are included in the F-CRH group. Tweets from these authors in F-CRH and OCN groups are used to analyze the second intervention between $\mathit{VisUS}$ and $\mathit{VisWorld}$. We perform one-to-one matching again for 976 tweets in the F-CRH group, and find 937 corresponding matches (96\%) from 83,505 tweets in the OCN group. Using matched tweets in both groups, we find that the coefficients of $\mathit{FCRH \times Vis}$ in Table \ref{tab:mis_out_did} are not significant across all columns. The extra reduction in engagement within the F-CRH group is no longer significant compared to the baseline before the interventions and relative to the OCN group. This further fails to support that the implementation of the Community Notes feature on X/Twitter had a statistically significant impact on the aggregated retweet count and like count in misleading tweets, thereby supporting the main claim of this paper. Moreover, in Table \ref{tab:mis_out_did}, the coefficients of $\mathit{Vis}$ and Month-Year fixed effects are not statistically significant, even in Columns (2) and (4). This indicates that the engagement of content on X/Twitter shows no significant variation for the same authors across different time intervals. It also implies that the decreases in retweet volume and likes are attributed to the inclusion of fact-checked tweets from authors who were newly included in each time period. Additionally, the results remain robust even under the shock of Musk's takeover.

In addition to the retweet count and like count, the engagement rate, which represents the ratio of retweets (retweet rate) or likes (like rate) to the total views, is another important metric for assessing the performance of Community Notes. This metric is utilized by X/Twitter itself to measure the effectiveness of the feature \cite{wojcik_birdwatch_2022}. Starting from December 15, 2022, X/Twitter has released a new public metric, impression count, to show the total number of times a Tweet has been viewed \cite{pfeffer_half-life_2023}. Because the impression count is only available after the implementation of Community Notes on the whole platform of X/Twitter, we cannot conduct the same regression analysis for retweet rate (like rate) as we do for retweet count (like count). Accordingly, we employ a binomial regression to investigate the association between Community Notes and engagement rate. The findings in Table \ref{tab:mis_out_did} demonstrate that controlling tweets from the same authors in different time periods can effectively mitigate the influence of confounding factors. In the analysis of the engagement rate, we focus on tweets from 515 authors who are included in the Community Notes during $\mathit{VisUS}$. Specifically, there are 328 tweets in the F-CRH group and 62,994 tweets in the OCN group after the inclusion of impression count as a public metric. Through one-to-one matching, we identify corresponding matches for 306 out of 328 tweets (93.3\%) in F-CRH (treatment group) from OCN (control group). We assume that the number of retweets (likes) for a given tweet $\mathit{i}$ follows a binomial distribution that is specified as \begin{equation}
    RetweetCount_{i} (LikeCount_{i}) \sim binomial(ImpressionCount_{i}, \theta_{i}),
\end{equation}
where \begin{math} \theta_{i} \in [0,1]\end{math} represents the probability that a user will retweet or like tweet $\mathit{i}$, and corresponds to the retweet (like) rate. We then use a $\mathit{logit}$ link function and specify the binomial regression model as:
\begin{equation}
    logit(\theta_{i}) = \beta_{0} + \beta_{1}Treated_{i} + \bm{\alpha^{'}x_{i}},
\end{equation}
where \begin{math} Treated_{i} = 1 \end{math} if tweet $\mathit{i}$ is in the treatment group (\ie, F-CRH group), otherwise \begin{math} = 0 \end{math}. $\mathit{\beta_{0}}$ is the intercept, and $\mathit{\bm{x_{i}}}$ denotes all the control variables described in Section \ref{sec:variable_extraction}.

As shown in Table \ref{tab:mis_out_bino_rate}, the coefficients of $\mathit{FCRH}$ in Columns (1) and (3) are statistically non-significant. It implies that the retweet rate and like rate are not significantly associated with Community Notes. This finding is robust under the control of Month-Year fixed effects in Columns (2) and (4). Additionally, we observe statistically significant negative coefficients for the variable $\mathit{Media}$ across all the columns in Table \ref{tab:mis_out_bino_rate}. Considering the positive effects of $\mathit{Media}$ on retweet and like counts in the previous findings, as shown in Table \ref{tab:mis_all_did}, we can interpret that tweets with media elements tend to have higher aggregated retweet (like) counts. However, it is important to note that these tweets also exhibit lower retweet (like) rates when compared to tweets without media elements.

\begin{table}
\caption{Regression results for retweet rate and like rate within F-CRH (treatment) and OCN (control) groups.}
\label{tab:mis_out_bino_rate}
\begin{tabular}{l*{4}{S}}
    \toprule
    &\multicolumn{2}{c}{Retweet Rate}&\multicolumn{2}{c}{Like Rate}\\
    \cmidrule(r){2-3} \cmidrule(r){4-5}
    &{\makecell{(1)}}&{\makecell{(2)}}&{\makecell{(3)}}&{\makecell{(4)}}\\
    \midrule
    $\mathit{FCRH}$&-0.014&-0.031&-0.388&-0.433\\
    &(0.249)&(0.263)&(0.256)&(0.253)\\
    $\mathit{WordNum}$&0.228^{**}&0.236^{**}&-0.058&-0.063\\
    &(0.083)&(0.081)&(0.076)&(0.075)\\
    $\mathit{Media}$&-0.616^{***}&-0.602^{***}&-0.529^{***}&-0.550^{***}\\
    &(0.136)&(0.147)&(0.122)&(0.136)\\
    $\mathit{Hashtag}$&-0.586&-0.480&-0.454&-0.268\\
    &(0.401)&(0.336)&(0.303)&(0.251)\\
    $\mathit{Mention}$&0.725&0.724&0.355&0.400\\
    &(0.404)&(0.404)&(0.329)&(0.331)\\
    $\mathit{AcctAge}$&-0.044&-0.040&-0.161^{**}&-0.161^{**}\\
    &(0.059)&(0.063)&(0.053)&(0.057)\\
    $\mathit{Verified}$&-0.184&-0.183&-0.096&-0.078\\
    &(0.250)&(0.264)&(0.219)&(0.232)\\
    $\mathit{Followers}$&-0.103&-0.103&-0.078^{**}&-0.085^{**}\\
    &(0.053)&(0.053)&(0.027)&(0.027)\\
    $\mathit{Followees}$&0.060&0.062&0.000&0.012\\
    &(0.062)&(0.063)&(0.049)&(0.051)\\
    MonthYear&\xmark&$\checkmark$&\xmark&$\checkmark$\\
    Emotions&$\checkmark$&$\checkmark$&$\checkmark$&$\checkmark$\\
    Topics&$\checkmark$&$\checkmark$&$\checkmark$&$\checkmark$\\
    $\mathit{Intercept}$&-6.285^{***}&-6.260^{***}&-4.343^{***}&-4.120^{***}\\
    &(0.347)&(0.404)&(0.341)&(0.382)\\
    Observations&{612}&{612}&{612}&{612}\\
    \bottomrule
    \multicolumn{5}{l}{\makecell[l]{Binomial regression models with robust standard errors in parentheses. \\ $^*p<0.05$, $^{**}p<0.01$, and $^{***}p<0.001$.}}\\
\end{tabular}
\end{table}

In sum, we conducted a thorough analysis on potential confounding factors stemming from the platform, contributors, tweet authors, and tweets themselves. The findings consistently fail to provide significant evidence supporting that the roll-out of Community Notes led to a reduction in engagement with misinformation in terms of aggregated retweet counts and likes. Furthermore, our analysis implies that the observed reduction in retweets and likes after the second intervention for both misleading and not misleading posts can be partially attributed to differences in the fact-checking targets, \ie, the incorporation of fact-checked tweets from authors who were newly included when the Community Notes feature was rolled out to the world.

\section{Discussion}
\label{sec:discussion}
The rise of social media provides unprecedented opportunities for individuals to share information and engage in public discourse. However, it has also created an environment where users can engage with misinformation easily, leading to negative impacts on individuals, society, and democracy as a whole. As such, engagement with online misinformation has become an important issue, and many interventions have been developed to address this problem. Evaluating the effectiveness of these interventions is crucial to ensure that efforts are directed towards the most promising strategies, and to inform future policy decisions. To our best knowledge, this study is the first to rigorously examine the effectiveness of crowdsourced fact-checking in reducing engagement with misinformation in the real world.

We use a longitudinal crowdsourced dataset of fact-checks from Community Notes to investigate whether this new feature effectively reduces engagement with misinformation on X/Twitter (as measured by retweet and like counts). Despite a significant increase in the number of notes and fact-checked tweets, we find that the expansion of Community Notes to the \US and the world might not have had a significant effect on reducing engagement with misinformation. This is confirmed by both DiD and RDD models, as well as comprehensive confounding analysis. Instead, our study reveals that the diffusion of both true and false fact-checked tweets significantly decreased in parallel when Community Notes became available to the world. Additionally, our results show that misleading tweets have lower virality compared to non-misleading tweets in terms of retweet and like counts, regardless of the expansion of the Community Notes feature.

\subsection{Research Implications}
Our analysis provides several important research implications. As our main contribution, we demonstrate that the roll-out of Community Notes had no statistically significant effect on reducing engagement with misinformation on X/Twitter in terms of retweet count and like count. This highlights the significance of utilizing real-world data for the evaluation of misinformation interventions \cite{courchesne_review_2021}. 

At a first glance, our findings seem to be inconsistent with the results of an A/B test carried out by X/Twitter during the pilot phase of Community Notes \cite{wojcik_birdwatch_2022}. X/Twitter found that users exposed to note annotations on tweets were 25--34\% less likely to like or retweet them compared to the control group. From this perspective, misleading tweets with CRH notes are expected to have lower engagement, compared to misleading tweets without CRH notes. However, this difference in engagement is not observed in our study. Importantly, however, this inconsistency does not necessarily mean that Community Notes cannot inform users and influence their engagement behaviors. Rather, a plausible explanation might be that the delay in note display to users may partially explain the insufficiency of Community Notes in reducing engagement with misinformation. While our subgroup analysis regarding different fact-checking delays fails to find supporting evidence for this notion, this may be because the smallest observed fact-checking delays in our data are still not short enough compared to the spread of source tweets. A recent study shows that about 95\% of tweets have no relevant impressions after two days, and it only takes about 79.5 minutes before half of the overall impressions are created for a tweet \cite{pfeffer_half-life_2023}. Similarly, research on Weibo, an X/Twitter-like service in China, suggests that approximately 80\% of reshares occur within the first two days of a post \cite{chuai_anger_2022}. Nevertheless, the results of our study reveal that, on average, fact-checking CRH notes require more than two days to appear directly on the corresponding source tweets, with the shortest response time observed being 80.2 minutes. This delay thus limits their ability to inform users at the early and most viral stage of misinformation diffusion.

Our current study primarily focuses on evaluating the aggregated engagement metrics (\ie, retweet count and like count) before and after the expansion of the Community Notes feature, rather than specifically examining the engagement timeline in each misleading tweet before and after the addition of notes. Given this limitation, we further consider potential confounding factors from multiple levels, including platform, contributors, authors, and tweets, to support our findings. Nonetheless, we also highlight the need for more detailed investigations to scrutinize the timelines and specific patterns of engagement change. Such fine-grained studies can provide deeper insights into the effects of Community Notes on user engagement. Theoretically, fact-checking annotations can reduce people's belief in misinformation \cite{wojcik_birdwatch_2022,clayton_real_2020}, but also may recall the backfire effect, which inadvertently strengthens misconceptions \cite{ecker_psychological_2022}. Moreover, previous research has indicated that users still engage with online misleading posts they distrust \cite{orosz_prosocial_2023,papakyriakopoulos_impact_2022,munyaka_misinformation_2022}. Additionally, despite existing deficits regarding critical reasoning and partisan bias in discerning falsehood, social media users also share information that challenges their own political ideologies \cite{ceylan_sharing_2023}. Since engaging with misinformation is part of broader complex social reactions \cite{ceylan_sharing_2023,epstein_social_2023}, future research should analyze the role of such patterns within crowd-based fact-checking in real-world social media environments.

We further observe that the engagement with misleading and not misleading fact-checked tweets decreased in parallel after Community Notes was expanded to the world. Several factors may contribute to this pattern. On the platform side, X/Twitter claims that they do not take any enforcement of other rules apart from the display of CRH notes on fact-checked tweets via Community Notes. Additionally, Musk's takeover and other time-specific confounding factors (such as real-world events) cannot explain the decline (see Appendix, Table \ref{tab:mis_all_did_ext}). From the aspect of community contributors, we find an enrollment of 48,927 new contributors (76\% of all contributors) in Community Notes after its expansion to the world, which is more than 24 times the number of enrolled contributors in the expansion limited to the \US With such a massive increase in the size of the community, unobserved changes may occur in the profiles of community contributors. However, differences between old contributors and new contributors are not statistically significant (see Appendix, Table \ref{tab:mis_all_did_ext}). Finally, when only comparing tweets from the same authors during different time periods, we find that the decline disappears, which means the decreases in retweet and like counts can be attributed to the inclusion of fact-checked tweets from authors who were newly included in each time period (see Appendix, Table \ref{tab:mis_out_did}). This also implies that the contributors, for both old and new enrolled ones, may focus on a wider range of tweets and not only the most viral ones after the pilot phase of Community Notes ended.

Ultimately, we find that community fact-checked misleading tweets consistently receive fewer retweets and likes than non-misleading tweets, regardless of whether the Community Notes program was expanded or not. A recent analysis of misinformation diffusion, which is based on the tweets fact-checked by Community Notes during its early pilot phase, also finds this phenomenon \cite{drolsbach_diffusion_2023}. Notably, this finding is contrary to the previous research that examines the spread of misinformation fact-checked by experts from third-party organizations \cite{vosoughi_spread_2018}. One potential explanation for this discrepancy is coming from the sample selection, \ie, that community fact-checkers may prioritize posts from influential accounts \cite{drolsbach_diffusion_2023,ma_characterizing_2023,pilarski_community_2023}. Additionally, our study also finds a positive correlation between verified accounts with a higher number of followers and receiving more notes and ratings. Nevertheless, the higher level of diffusion in not misleading tweets still exists under the control of the variables that denote the influence of authors (the coefficients of $\mathit{FCRH}$ in Table \ref{tab:tf_high_rank_did}), which calls for more detailed explanations.

\subsection{Practical Implications}
This study also provides important practical implications for harnessing the ``wisdom of crowd'' as a countermeasure against misinformation on social media platforms. First, the significant increase of individual fact-checkers in Community Notes after its implementation on X/Twitter makes it feasible to fact-check online information at large scale. Moreover, through active individual fact-checkers on social media, specific pieces of misinformation that come from highly influential accounts or raise significant concerns can be checked directly at their original places. 

Second, while crowdsourced fact-checking, such as Community Notes, can provide a scalable way to detect and label misinformation \cite{drolsbach_diffusion_2023, allen_scaling_2021}, they may not lead to significant reductions in the engagement with misinformation. The lack of a significant effect of Community Notes on mitigating overall engagement with misinformation should remind social media platforms to reconsider their crowdsourced fact-checking strategies. Reliable verdicts require sufficient assessments from different perspectives, which means that corresponding pieces of misinformation must first gain attention and become widespread to receive enough ratings. To address this post-diffusion dilemma with crowdsourced fact-checking, social media platforms could implement a pre-selection mechanism to filter tweets with a high probability of being misleading at the beginning of their diffusion and then selectively push them to community fact-checkers. By incentivizing contributors to act promptly, such a pre-selection mechanism could reduce delays in obtaining fact-checking verdicts. Notably, while Community Notes may not have directly reduced the overall engagement with misinformation, it may still provide a valuable feature to keep users well-informed with helpful context.

Third, our study also highlights the importance of measuring the impact of fact-checking interventions using rigorous methods and real-world data, rather than relying solely on controlled experiments or simulations. By analyzing large-scale datasets of user behaviors and social interactions, researchers can gain a better understanding of the complex dynamics of misinformation propagation and evaluate the effectiveness of various interventions. 

Finally, our study also underscores the importance of improving media literacy among users. Social reactions, such as retweet and like, are the essential components that make social media ``social'' \cite{epstein_social_2023}. There are complex interaction mechanisms among online users. While fact-checking can be an effective tool for identifying and flagging false claims, it may not be sufficient to reduce engagement with misinformation. Therefore, social media platforms, governments, and public organizations should work together to promote critical thinking, digital literacy, and transparency in online communication \cite{saltz_misinformation_2021}.

\subsection{Limitations and Future Research}
There are several limitations in our study. First, because the dataset is mainly in English and the program is well-established in the \US, we conduct the research with tweets only in English. Multilingual evaluation should be examined with the establishment of Community Notes to get a more comprehensive understanding. Second, we focus on the effect of Community Notes on the engagement with misleading tweets and use the T-NMR group as a mediating group indicating non-misleading information. However, tweets in the T-NMR group also have the potential to be misleading, because of general misconception or manipulation \cite{drolsbach_diffusion_2023,allen_birds_2022,bhuiyan_investigating_2020}. To mitigate this concern, we implement rigorous selection criteria to ensure that only tweets with a high likelihood of being non-misleading are included in the T-NMR group. Relatedly, F-NCRH tweets serve as the control group for comparison with the F-CRH group. However, certain F-NCRH notes may not be rated as helpful because the corresponding source tweets are not misleading (\eg, due to misuse of the fact-checking feature). To account for this, we exclude T-NMR tweets from F-NCRH tweets, specifically those containing both T-NMR notes and F-NCRH notes, and use propensity score matching to filter the F-NCRH group to enhance its comparability with the F-CRH group to the largest extent possible. Third, 14\% of the total tweets fact-checked via Community Notes are unavailable via the API (see Section \ref{sec:data_collection}) and thus cannot be included in the evaluation. Some of these unavailable tweets might be misleading and removed by the authors when the CRH notes appeared. Additionally, some particularly egregious misinformation might have been removed by the platform (\ie, X/Twitter). While the retrieval rate for tweets with CRH notes is similar to the overall rate for all tweets, this could potentially still lead to an underestimation of the effectiveness of Community Notes. Fourth, our study is limited to X/Twitter, and the findings may not be generalizable to other social media platforms with different user demographics and content characteristics. Finally, given the challenges of accessing detailed engagement timelines from X/Twitter, we had to rely on aggregated retweet count and like count as metrics to analyze the effectiveness of Community Notes. Future research should try to further explore the effect of Community Notes and other misinformation interventions with more fine-grained social media data. This would provide a more comprehensive understanding of the effects and mechanisms underlying these interventions, potentially leading to more effective strategies in reducing engagement with misinformation on social media platforms.

\section{Conclusion}

In this paper, we conduct a large-scale empirical study to evaluate the effectiveness of X/Twitter's Community Notes feature in reducing engagement with misinformation on X/Twitter. Based on the results from both DiD and RDD models, we find no evidence that the introduction of Community Notes and its roll-out to users in the \US and around the world led to a significant reduction in user engagement with misleading tweets. Our findings suggest that the response time of Community Notes may be not fast enough to intervene in the engagement with misinformation during early and highly viral stages of diffusion. Taken together, our work emphasizes the importance of evaluating fact-checking interventions in the field and provides important implications for improving crowdsourced fact-checking strategies on social media platforms.

\section{Ethical Statement}
This research has received ethical approval from the Ethics Review Panel of the University of Luxembourg (ref. ERP 23-053 REMEDIS). 
All analyses are based on publicly available data. To respect privacy, we explicitly do not publish usernames in our paper and only report aggregate results. We declare no competing interests.

\begin{acks}
This research is supported by the Luxembourg National Research Fund (FNR) and Belgian National Fund for Scientific Research (FNRS), as part of the project REgulatory Solutions to MitigatE DISinformation (REMEDIS), grant ref. INTER\_FNRS\_21\_16554939\_REMEDIS. Furthermore, this research is supported by a research grant from the German Research Foundation (DFG grant 492310022). 
\end{acks}

\bibliographystyle{ACM-Reference-Format-no-doi-abbrv}
\bibliography{refs}

\appendix
\section*{Appendix}
Fig. \ref{fig:did_rdd_groups}-\ref{fig:ccdf_tf_phases} and Table \ref{tab:domains2topics}-\ref{tab:parallel_tf_like}.
\clearpage
\renewcommand\thetable{A\arabic{table}}
\setcounter{table}{0}
\renewcommand\thefigure{A\arabic{figure}}
\setcounter{figure}{0}


\begin{figure*}
    \centering
    \subfigure[DiD]{
    \begin{minipage}{0.45\linewidth}
        \centering
        \includegraphics[width=\columnwidth]{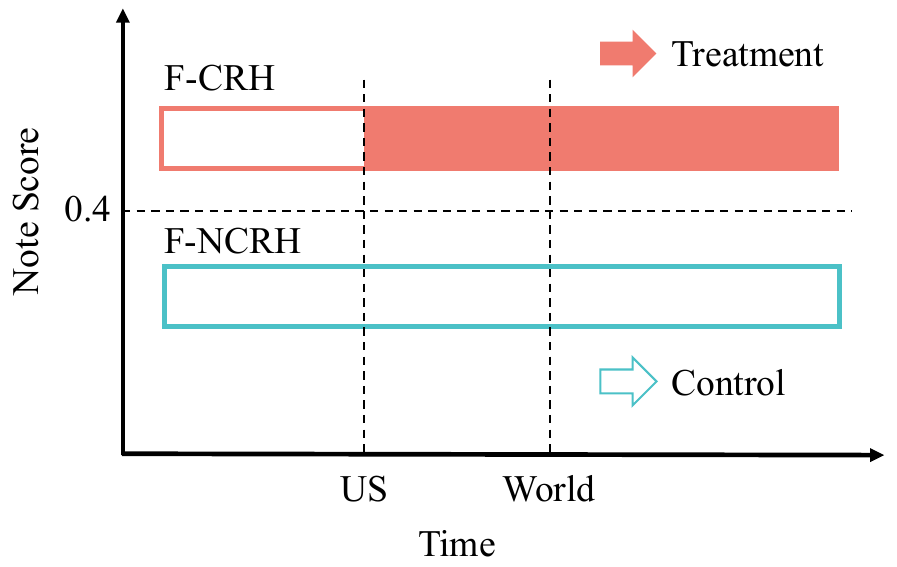}
    \end{minipage}
    }
    \subfigure[RDD]{
    \begin{minipage}{0.45\linewidth}
        \centering
        \includegraphics[width=\columnwidth]{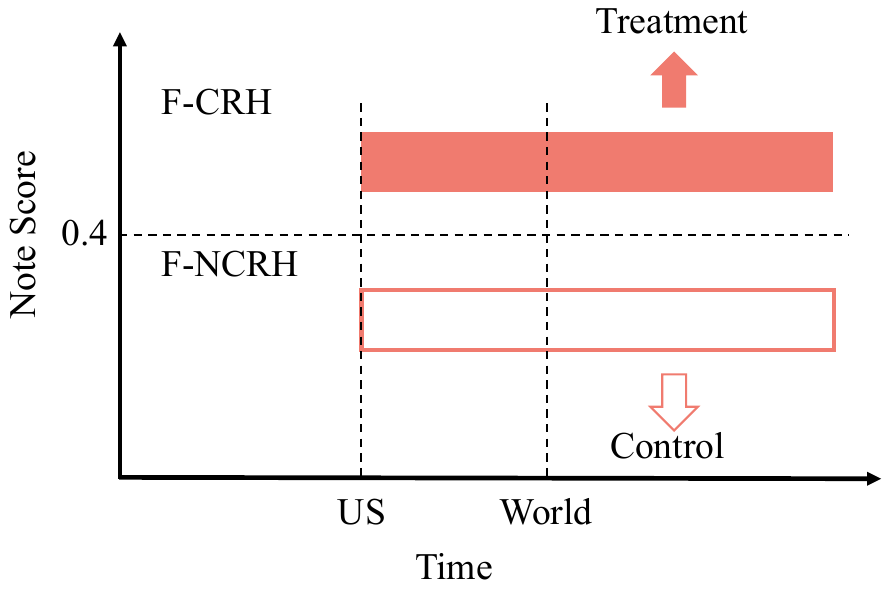}
    \end{minipage}
    }
    \caption{The illustrations of F-CRH (treatment) and F-NCRH (control) groups in (a) DiD and (b) RDD models.}
    \label{fig:did_rdd_groups}
\end{figure*}

\clearpage
\begin{figure}
    \centering
    \subfigure[Parallel test for F-CRH (treatment) and F-NCRH (control) groups.]{
    \begin{minipage}{\linewidth}
        \centering
        \includegraphics[scale=0.6]{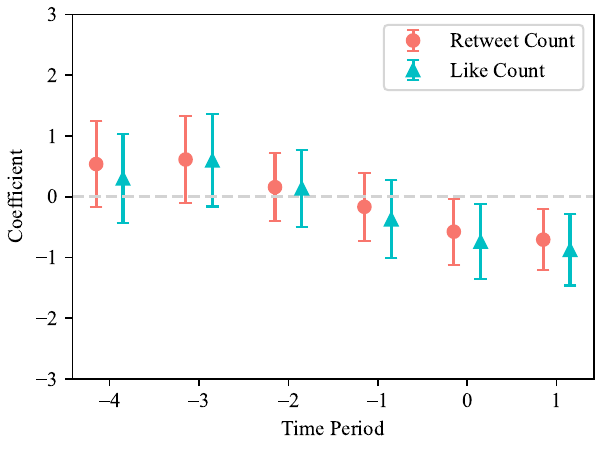}
    \end{minipage}
    }

    \subfigure[Parallel test for T-NMR (treatment) and F-NCRH (control) groups.]{
    \begin{minipage}{\linewidth}
        \centering
        \includegraphics[scale=0.6]{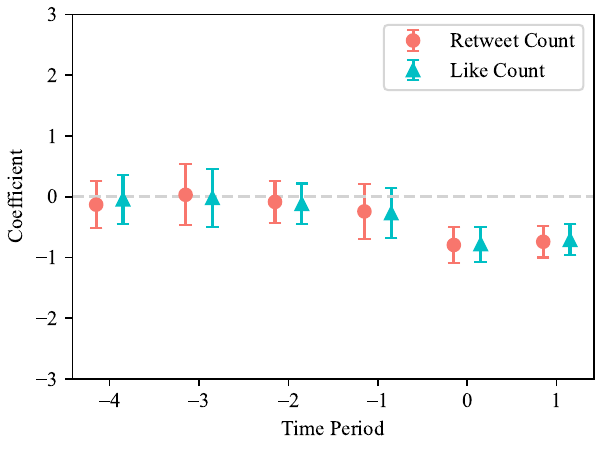}
    \end{minipage}
    }
    
    \subfigure[Parallel test for F-CRH (treatment) and T-NMR (control) groups.]{
    \begin{minipage}{\linewidth}
        \centering
        \includegraphics[scale=0.6]{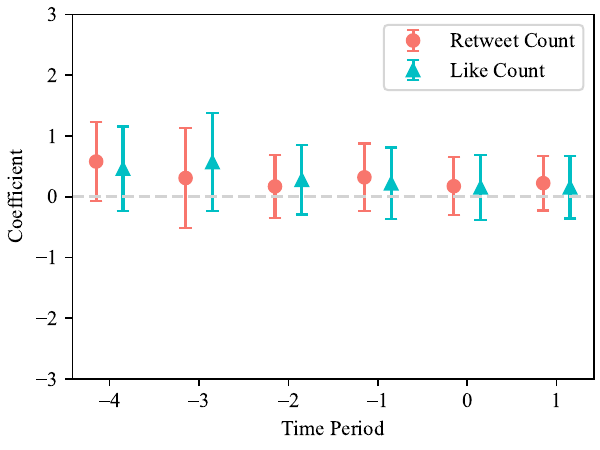}
    \end{minipage}
    }
    \caption{Parallel tests of retweet count and like count between the treatment and control groups. The baseline is the first three-month period spanning from July 1, 2021 to September 30, 2021. (a) Parallel test for F-CRH (treatment) and F-NCRH (control) groups. The regression results are shown in Table \ref{tab:parallel_mis_did_retweet} and \ref{tab:parallel_mis_did_like}. (b) Parallel test for T-NMR (treatment) and F-NCRH (control) groups. The regression results are shown in Table \ref{tab:parallel_mis_did_retweet_placebo} and \ref{tab:parallel_mis_did_like_placebo}. (c) Parallel test for F-CRH (treatment) and T-NMR (control) groups. The regression results are shown in Table \ref{tab:parallel_tf_retweet} and \ref{tab:parallel_tf_like}.}
    \label{fig:parallel_test}
\end{figure}

\clearpage
\begin{figure}
    \centering
    \subfigure[CCDFs of retweet count for misleading and not misleading tweets.]{
    \begin{minipage}{\linewidth}
        \centering
        \includegraphics[scale=0.7]{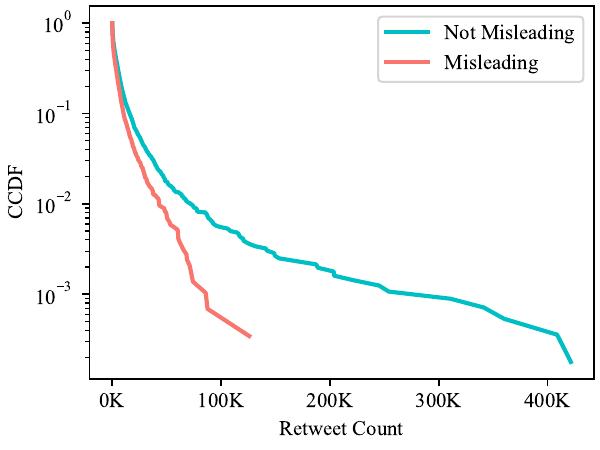}
    \end{minipage}
    }
    
    \subfigure[CCDFs of like count for misleading and not misleading tweets.]{
    \begin{minipage}{\linewidth}
        \centering
        \includegraphics[scale=0.7]{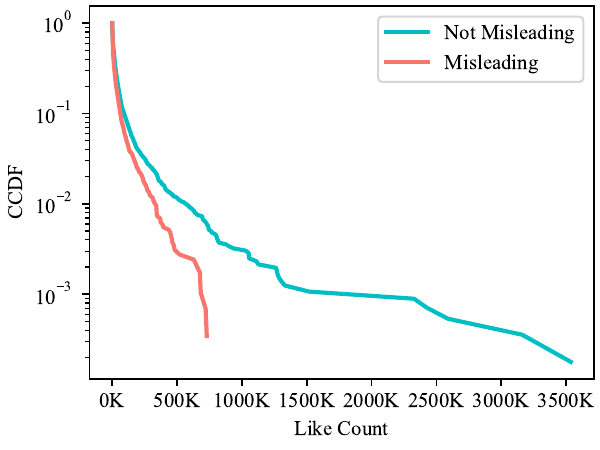}
    \end{minipage}
    }
    \caption{CCDFs of retweet count and like count based on the fact-checked tweets from matched F-CRH (Misleading) and T-NMR (Not Misleading) groups. (a) CCDFs of retweet count for misleading and not misleading tweets. $KS=0.121, p<0.001$. (b) CCDFs of like count for misleading and not misleading tweets. $KS=0.133, p<0.001$.}
    \label{fig:ccdf_tf}
\end{figure}

\clearpage
\begin{figure}
    \centering
    \subfigure[Before 2022-10-06]{
    \begin{minipage}{0.45\linewidth}
        \centering
        \includegraphics[width=\columnwidth]{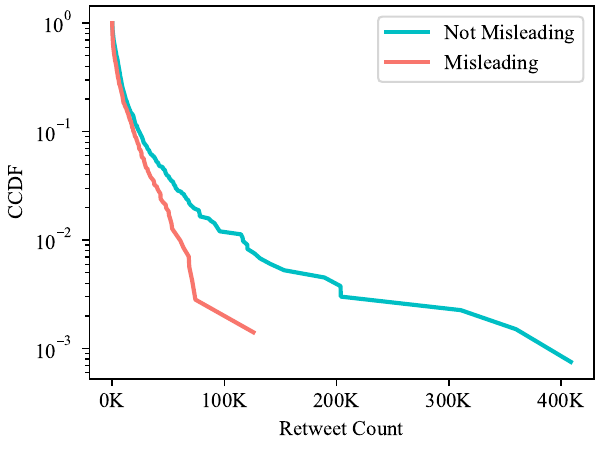}
    \end{minipage}
    }
    \subfigure[Before 2022-10-06]{
    \begin{minipage}{0.45\linewidth}
        \centering
        \includegraphics[width=\columnwidth]{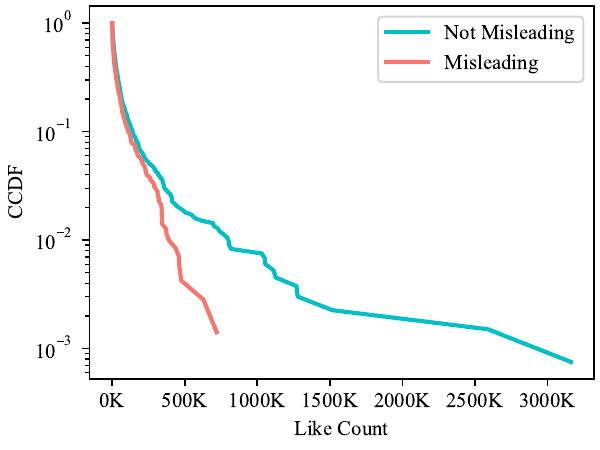}
    \end{minipage}
    }
    
    \subfigure[Between 2022-10-06 and 2022-12-11 (US)]{
    \begin{minipage}{0.45\linewidth}
        \centering
        \includegraphics[width=\columnwidth]{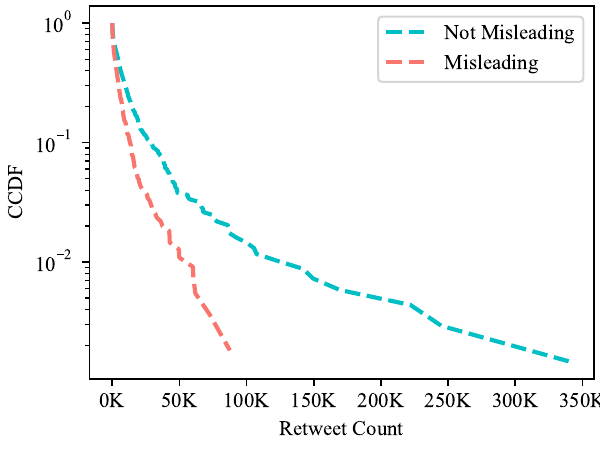}
    \end{minipage}
    }
    \subfigure[Between 2022-10-06 and 2022-12-11 (US)]{
    \begin{minipage}{0.45\linewidth}
        \centering
        \includegraphics[width=\columnwidth]{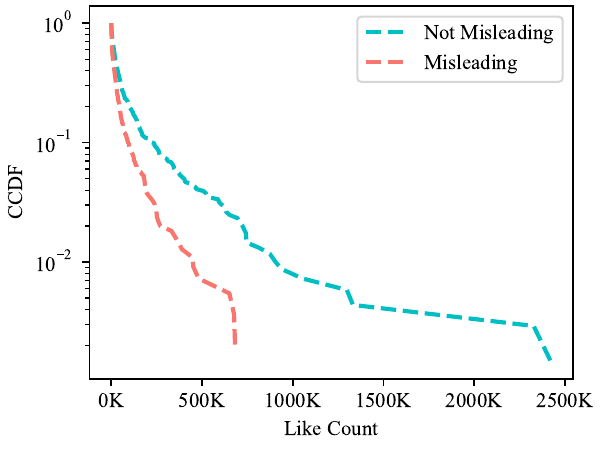}
    \end{minipage}
    }

    \subfigure[After 2022-12-11 (World)]{
    \begin{minipage}{0.45\linewidth}
        \centering
        \includegraphics[width=\columnwidth]{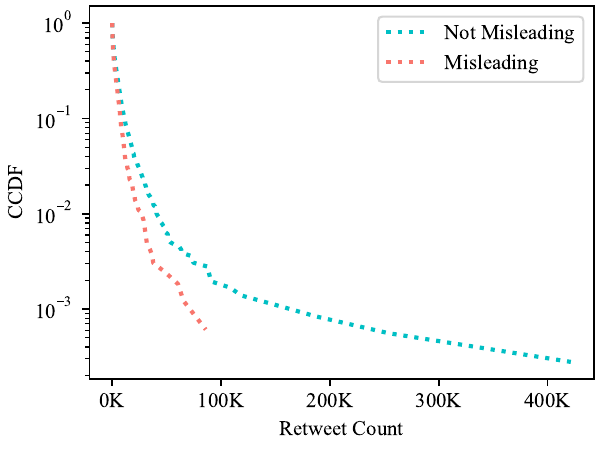}
    \end{minipage}
    }
    \subfigure[After 2022-12-11 (World)]{
    \begin{minipage}{0.45\linewidth}
        \centering
        \includegraphics[width=\columnwidth]{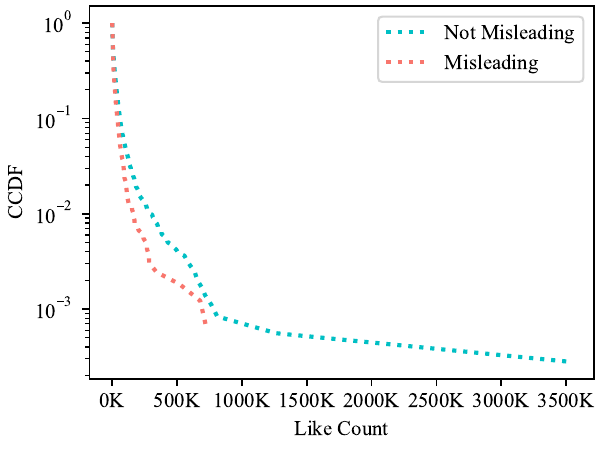}
    \end{minipage}
    }
    \caption{CCDFs of retweet count and like count based on the fact-checked tweets from matched F-CRH (Misleading) and T-NMR (Not Misleading) groups, during different time periods. (a) CCDFs of retweet count for misleading and not misleading tweets before 2022-10-06. $KS=0.153, p<0.001$. (b) CCDFs of like count for misleading and not misleading tweets before 2022-10-06. $KS=0.168, p<0.001$. (c) CCDFs of retweet count for misleading and not misleading tweets between 2022-10-06 and 2022-12-11 (US). $KS=0.186, p<0.001$. (d) CCDFs of like count for misleading and not misleading tweets between 2022-10-06 and 2022-12-11 (US). $KS=0.205, p<0.001$. (e) CCDFs of retweet count for misleading and not misleading tweets after 2022-12-11 (World). $KS=0.128, p<0.001$. (f) CCDFs of like count for misleading and not misleading tweets after 2022-12-11 (World). $KS=0.133, p<0.001$.}
    \label{fig:ccdf_tf_phases}
\end{figure}


\clearpage
\begin{table}
  \caption{Topic list with corresponding domains}
  \label{tab:domains2topics}
  \resizebox{\textwidth}{!}{
  \begin{tabular}{ccl}
    \toprule
    Topic & Number & Domain\\
    \midrule
    Entertainment&54&\makecell[l]{Actor, American Football Game, Athlete, Award Show, Baseball Game, Basketball Game, \\ Book, Book Genre, Coach, Concert, Cricket Match, Entertainment Personality, \\ Esports League, Esports Player, Esports Team, Exercise \& Fitness, Fan Community, \\ Fictional Character, Global TV Show, Hockey Game, Holiday, Interests and Hobbies, \\ Interests and Hobbies Category, Interests and Hobbies Vertical, Movie, Movie Festival, \\ Movie Genre, Multimedia Franchise, Music Album, Music Genre, Musician, NFL Football Game, \\ Podcast, Points of Interest, Radio Station, Soccer Match, Sport, Sports Event, \\ Sports League, Sports Personality, Sports Series, Sports Team, TV Channels, \\ TV Episodes, TV Genres, TV Shows, Travel, Video Game, Video Game Conference, Song, \\ Video Game Hardware, Video Game Personality, Video Game Publisher, Video Game Tournament}\\
    \\
    Finance \& Business&8& \makecell[l]{Brand, Brand Vertical, Business Taxonomy, Digital Assets \& Crypto, Product, \\ Product Taxonomy, Product Version, Stocks}\\
    \\
    Politics&3&Political Body, Political Race, Politician\\
    \\
    Science \& Technology&3&Fields of study, Technology, Video Game Hardware\\
    \\
    Emergency Events&1&Emergency Events\\
    \\
    Ongoing News Story&1&Ongoing News Story\\
    \bottomrule
  \end{tabular}}
\end{table}

\clearpage
\begin{table}
\caption{Descriptive Statistics}
\label{tab:desc_vars}
\resizebox{\textwidth}{!}{
\begin{tabular}{lcccccccc}
    \toprule
    {Variable}&{Count}&{Mean}&{Std}&{Min}&{25\%}&{50\%}&{75\%}&{Max}\\
    \midrule
    $RetweetCount$&30267&2865.39&9402.57&0&27&346&2336.5&421330\\
    $LikeCount$&30267&16470.3&66307.5&0&125&1544&10447&3.53659e+06\\
    $\mathit{WordNum}$&30267&15.45&7.923&0&9&15&22&81\\
    $\mathit{Media}$&30267&0.408&0.492&0&0&0&1&1\\
    $\mathit{Hashtag}$&30267&0.098&0.298&0&0&0&0&1\\
    $\mathit{Mention}$&30267&0.209&0.406&0&0&0&0&1\\
    $\mathit{AcctAge}$&30267&3093.3&1790.13&0&1300.5&3577&4711&19305\\
    $\mathit{Verified}$&30267&0.674&0.469&0&0&1&1&1\\
    $\mathit{Followers}$&30267&3.56078e+06&1.65064e+07&0&14589&141159&746093&1.34151e+08\\
    $\mathit{Followees}$&30267&8143.63&39808.2&0&388.5&1176&3444.5&610998\\
    $\mathit{Anger}$&30267&0.148&0.184&0&0&0.111&0.25&1\\
    $\mathit{Disgust}$&30267&0.078&0.142&0&0&0&0.143&1\\
    $\mathit{Joy}$&30267&0.131&0.246&0&0&0&0.167&1\\
    $\mathit{Fear}$&30267&0.213&0.24&0&0&0.2&0.333&1\\
    $\mathit{Sadness}$&30267&0.129&0.174&0&0&0&0.222&1\\
    $\mathit{Surprise}$&30267&0.108&0.196&0&0&0&0.167&1\\
    $\mathit{EmergencyEvents}$&30267&0.001&0.024&0&0&0&0&1\\
    $\mathit{Entertainment}$&30267&0.236&0.424&0&0&0&0&1\\
    $\mathit{FinanceBusiness}$&30267&0.259&0.438&0&0&0&1&1\\
    $\mathit{OngoingNewsStory}$&30267&0.143&0.35&0&0&0&0&1\\
    $\mathit{Politics}$&30267&0.336&0.472&0&0&0&1&1\\
    $\mathit{ScienceTechnology}$&30267&0.127&0.334&0&0&0&0&1\\
    \bottomrule
\end{tabular}}
\end{table}

\clearpage
\begin{table}
\caption{DiD regression results for retweet count and like count within F-CRH (treatment) and F-NCRH (control) groups between Pilot and Roll-Out periods.}
\label{tab:mis_all_did_rollout}
\begin{tabular}{l*{2}{S}}
    \toprule
    &{(1)}&{(2)}\\
    &\multicolumn{1}{c}{Retweet Count}&\multicolumn{1}{c}{Like Count}\\
    \midrule
    $\mathit{FCRH}$&1.236^{***}&1.282^{***}\\
    &(0.144)&(0.152)\\
    $\mathit{Vis}$&0.145^{}&0.152^{}\\
    &(0.080)&(0.090)\\
    $\mathit{FCRH \times Vis}$&-0.487^{**}&-0.455^{**}\\
    &(0.152)&(0.161)\\
    $\mathit{WordNum}$&0.141^{***}&-0.015^{}\\
    &(0.028)&(0.030)\\
    $\mathit{Media}$&0.668^{***}&0.512^{***}\\
    &(0.053)&(0.058)\\
    $\mathit{Hashtag}$&-0.742^{***}&-0.986^{***}\\
    &(0.089)&(0.090)\\
    $\mathit{Mention}$&-1.301^{***}&-1.323^{***}\\
    &(0.066)&(0.069)\\
    $\mathit{AcctAge}$&-0.017^{}&0.009^{}\\
    &(0.023)&(0.025)\\
    $\mathit{Verified}$&0.764^{***}&0.703^{***}\\
    &(0.056)&(0.059)\\
    $\mathit{Followers}$&0.180^{***}&0.223^{***}\\
    &(0.013)&(0.012)\\
    $\mathit{Followees}$&0.029^{}&0.031^{**}\\
    &(0.016)&(0.012)\\
    $\mathit{VisWorld}$&-0.782^{***}&-0.834^{***}\\
    &(0.057)&(0.061)\\
    Emotions&$\checkmark$&$\checkmark$\\
    Topics&$\checkmark$&$\checkmark$\\
    $\mathit{Intercept}$&6.847^{***}&8.611^{***}\\
    &(0.082)&(0.095)\\
    Observations&{13,808}&{13,808}\\
    \bottomrule
    \multicolumn{3}{l}{\makecell[l]{Negative binomial regression models with robust standard errors \\ in parentheses. $^*p<0.05$, $^{**}p<0.01$, and $^{***}p<0.001$.}}\\
\end{tabular}
\end{table}

\begin{table}
\caption{DiD regression results for retweet count and like count within T-NMR (treatment) and F-NCRH (control) groups.}
\label{tab:placebo_test_did}
\begin{tabular}{l*{4}{S}}
    \toprule
    &\multicolumn{2}{c}{Retweet Count}&\multicolumn{2}{c}{Like Count}\\
    \cmidrule(r){2-3} \cmidrule(r){4-5}
    &{\makecell{(1)\\Pre-US}}&{\makecell{(2)\\US-World}}&{\makecell{(3)\\Pre-US}}&{\makecell{(4)\\US-World}}\\
    \midrule
    $\mathit{TNMR}$&1.565^{***}&1.026^{***}&1.569^{***}&1.086^{***}\\
    &(0.185)&(0.096)&(0.166)&(0.092)\\
    $\mathit{Vis}$&0.166^{*}&-0.765^{***}&0.190^{*}&-0.816^{***}\\
    &(0.081)&(0.064)&(0.086)&(0.070)\\
    $\mathit{TNMR \times Vis}$&-0.503^{*}&0.060&-0.439^{*}&0.078\\
    &(0.204)&(0.112)&(0.189)&(0.111)\\
    $\mathit{WordNum}$&0.106^{**}&0.123^{***}&-0.053&-0.042\\
    &(0.038)&(0.029)&(0.041)&(0.030)\\
    $\mathit{Media}$&0.477^{***}&0.639^{***}&0.366^{***}&0.409^{***}\\
    &(0.081)&(0.054)&(0.085)&(0.058)\\
    $\mathit{Hashtag}$&-0.686^{***}&-0.618^{***}&-0.884^{***}&-0.851^{***}\\
    &(0.149)&(0.095)&(0.142)&(0.091)\\
    $\mathit{Mention}$&-1.249^{***}&-1.183^{***}&-1.218^{***}&-1.161^{***}\\
    &(0.099)&(0.065)&(0.108)&(0.067)\\
    $\mathit{AcctAge}$&-0.026&-0.052^{*}&-0.001&-0.027\\
    &(0.032)&(0.024)&(0.031)&(0.026)\\
    $\mathit{Verified}$&0.693^{***}&0.858^{***}&0.560^{***}&0.775^{***}\\
    &(0.086)&(0.061)&(0.086)&(0.064)\\
    $\mathit{Followers}$&0.233^{***}&0.235^{***}&0.286^{***}&0.291^{***}\\
    &(0.021)&(0.016)&(0.020)&(0.015)\\
    $\mathit{Followees}$&0.045&0.017&0.032&0.021\\
    &(0.047)&(0.016)&(0.051)&(0.012)\\
    Emotions&$\checkmark$&$\checkmark$&$\checkmark$&$\checkmark$\\
    Topics&$\checkmark$&$\checkmark$&$\checkmark$&$\checkmark$\\
    $\mathit{Intercept}$&7.048^{***}&6.892^{***}&8.837^{***}&8.706^{***}\\
    &(0.100)&(0.078)&(0.109)&(0.086)\\
    Observations&{5,044}&{13,328}&{5,044}&{13,328}\\
    \bottomrule
    \multicolumn{5}{l}{\makecell[l]{Negative binomial regression models with robust standard errors \\ in parentheses. $^*p<0.05$, $^{**}p<0.01$, and $^{***}p<0.001$.}}\\
\end{tabular}
\end{table}

\clearpage
\begin{table}
  \caption{DiD regression results for retweet count and like count within F-CRH (treatment) and T-NMR (control) groups under the limitation of the top 25\% delay in note display.}
  \label{tab:tf_high_rank_did_delay}
  \begin{tabular}{l*{4}{S}}
    \toprule
    &\multicolumn{2}{c}{Retweet Count}&\multicolumn{2}{c}{Like Count}\\
    \cmidrule(r){2-3} \cmidrule(r){4-5}
    &{\makecell{(1)\\Pre-US}}&{\makecell{(2)\\US-World}}&{\makecell{(3)\\Pre-US}}&{\makecell{(4)\\US-World}}\\
    \midrule
    $\mathit{FCRH}$&-0.267&-0.343^{*}&-0.289&-0.184\\
    &(0.154)&(0.168)&(0.149)&(0.173)\\
    $\mathit{Vis}$&-0.190&-0.684^{***}&-0.120&-0.749^{***}\\
    &(0.130)&(0.084)&(0.128)&(0.079)\\
    $\mathit{FCRH \times Vis}$&-0.009&-0.141&0.113&-0.261\\
    &(0.216)&(0.188)&(0.215)&(0.198)\\
    $\mathit{WordNum}$&-0.036&0.058&-0.216^{***}&-0.132^{***}\\
    &(0.051)&(0.032)&(0.051)&(0.033)\\
    $\mathit{Media}$&-0.217^{*}&0.314^{***}&-0.245^{*}&0.096\\
    &(0.105)&(0.065)&(0.109)&(0.069)\\
    $\mathit{Hashtag}$&-0.055&-0.390^{**}&-0.305&-0.598^{***}\\
    &(0.287)&(0.132)&(0.242)&(0.128)\\
    $\mathit{Mention}$&-0.612^{***}&-0.520^{***}&-0.675^{***}&-0.525^{***}\\
    &(0.128)&(0.081)&(0.133)&(0.080)\\
    $\mathit{AcctAge}$&-0.039&-0.109^{**}&0.011&-0.114^{***}\\
    &(0.044)&(0.033)&(0.045)&(0.034)\\
    $\mathit{Verified}$&-0.023&0.351^{***}&-0.113&0.285^{***}\\
    &(0.130)&(0.075)&(0.125)&(0.077)\\
    $\mathit{Followers}$&0.255^{***}&0.295^{***}&0.279^{***}&0.327^{***}\\
    &(0.032)&(0.023)&(0.030)&(0.020)\\
    $\mathit{Followees}$&0.008&-0.023&0.036&-0.009\\
    &(0.080)&(0.012)&(0.084)&(0.012)\\
    Emotions&$\checkmark$&$\checkmark$&$\checkmark$&$\checkmark$\\
    Topics&$\checkmark$&$\checkmark$&$\checkmark$&$\checkmark$\\
    $\mathit{Intercept}$&9.182^{***}&8.401^{***}&10.952^{***}&10.320^{***}\\
    &(0.178)&(0.109)&(0.166)&(0.105)\\
    Observations&{1,293}&{4,823}&{1,716}&{6,463}\\
    \bottomrule
    \multicolumn{5}{l}{\makecell[l]{Negative binomial regression models with robust standard errors \\ in parentheses. $^*p<0.05$, $^{**}p<0.01$, and $^{***}p<0.001$.}}\\
\end{tabular}
\end{table}

\clearpage
\begin{table}
\caption{DiD regression results for retweet count and like count between Pilot and Roll-Out periods in placebo tests.}
\label{tab:placebo_rollout}
\begin{tabular}{l*{4}{S}}
    \toprule
    &\multicolumn{2}{c}{T-NMR -- F-NCRH}&\multicolumn{2}{c}{F-CRH -- T-NMR}\\
    \cmidrule(r){2-3} \cmidrule(r){4-5}
    &{(1)}&{(2)}&{(3)}&{(4)}\\
    &\multicolumn{1}{c}{Retweet Count}&\multicolumn{1}{c}{Like Count}&\multicolumn{1}{c}{Retweet Count}&\multicolumn{1}{c}{Like Count}\\
    \midrule
    $\mathit{TNMR}$&1.640^{***}&1.623^{***}&&\\
    &(0.217)&(0.188)&&\\
    $\mathit{FCRH}$&&&-0.315&-0.316\\
    &&&(0.187)&(0.173)\\
    $\mathit{Vis}$&0.122&0.134&-0.265&-0.181\\
    &(0.079)&(0.091)&(0.157)&(0.143)\\
    $\mathit{TNMR \times Vis}$&-0.557^{*}&-0.464^{*}&&\\
    &(0.223)&(0.196)&&\\
    $\mathit{FCRH \times Vis}$&&&-0.079&-0.037\\
    &&&(0.195)&(0.186)\\
    $\mathit{WordNum}$&0.123^{***}&-0.041&0.028&-0.141^{***}\\
    &(0.026)&(0.027)&(0.028)&(0.032)\\
    $\mathit{Media}$&0.617^{***}&0.411^{***}&0.239^{***}&0.095\\
    &(0.050)&(0.054)&(0.055)&(0.061)\\
    $\mathit{Hashtag}$&-0.644^{***}&-0.869^{***}&-0.495^{***}&-0.678^{***}\\
    &(0.088)&(0.085)&(0.112)&(0.116)\\
    $\mathit{Mention}$&-1.194^{***}&-1.189^{***}&-0.594^{***}&-0.642^{***}\\
    &(0.060)&(0.062)&(0.071)&(0.071)\\
    $\mathit{AcctAge}$&-0.049^{*}&-0.023&-0.089^{***}&-0.076^{**}\\
    &(0.021)&(0.023)&(0.026)&(0.027)\\
    $\mathit{Verified}$&0.792^{***}&0.689^{***}&0.207^{***}&0.198^{**}\\
    &(0.055)&(0.059)&(0.061)&(0.064)\\
    $\mathit{Followers}$&0.237^{***}&0.296^{***}&0.292^{***}&0.328^{***}\\
    &(0.015)&(0.014)&(0.022)&(0.019)\\
    $\mathit{Followees}$&0.019&0.022&-0.012&0.005\\
    &(0.016)&(0.012)&(0.012)&(0.013)\\
    $\mathit{VisWorld}$&-0.747^{***}&-0.796^{***}&-0.643^{***}&-0.727^{***}\\
    &(0.054)&(0.058)&(0.062)&(0.063)\\
    Emotions&$\checkmark$&$\checkmark$&$\checkmark$&$\checkmark$\\
    Topics&$\checkmark$&$\checkmark$&$\checkmark$&$\checkmark$\\
    $\mathit{Intercept}$&6.847^{***}&8.653^{***}&8.837^{***}&10.603^{***}\\
    &(0.081)&(0.095)&(0.163)&(0.145)\\
    Observations&{15,354}&{15,354}&{6,944}&{6,944}\\
    \bottomrule
    \multicolumn{5}{l}{\makecell[l]{Negative binomial regression models with robust standard errors \\ in parentheses. $^*p<0.05$, $^{**}p<0.01$, and $^{***}p<0.001$.}}\\
\end{tabular}
\end{table}

\clearpage
\begin{table}
\caption{DiD regression results with external controls for retweet count and like count within F-CRH (treatment) and F-NCRH (control) groups.}
\label{tab:mis_all_did_ext}
\resizebox{0.6\textwidth}{!}{
\begin{tabular}{l*{4}{S}}
    \toprule
    &\multicolumn{2}{c}{Retweet Count}&\multicolumn{2}{c}{Like Count}\\
    \cmidrule(r){2-3} \cmidrule(r){4-5}
    &{\makecell{(1)\\Pre-US}}&{\makecell{(2)\\US-World}}&{\makecell{(3)\\Pre-US}}&{\makecell{(4)\\US-World}}\\
    \midrule
    $\mathit{FCRH}$&1.172^{***}&0.860^{***}&1.224^{***}&0.947^{***}\\
    &(0.134)&(0.117)&(0.140)&(0.117)\\
    $\mathit{Vis}$&0.182&0.019&0.121&0.198\\
    &(0.304)&(0.125)&(0.310)&(0.136)\\
    $\mathit{FCRH \times Vis}$&-0.438^{**}&-0.102&-0.411^{*}&-0.083\\
    &(0.163)&(0.132)&(0.168)&(0.135)\\
    $\mathit{Musk}$&0.097&0.165&0.193&0.213\\
    &(0.199)&(0.260)&(0.187)&(0.239)\\
    $\mathit{OldWriter}$&0.026&-0.110&0.033&-0.026\\
    &(0.091)&(0.070)&(0.093)&(0.070)\\
    $\mathit{WordNum}$&0.082^{*}&0.139^{***}&-0.072&-0.011\\
    &(0.037)&(0.031)&(0.041)&(0.032)\\
    $\mathit{Media}$&0.510^{***}&0.763^{***}&0.454^{***}&0.598^{***}\\
    &(0.076)&(0.059)&(0.083)&(0.059)\\
    $\mathit{Hashtag}$&-0.870^{***}&-0.707^{***}&-1.059^{***}&-0.896^{***}\\
    &(0.150)&(0.091)&(0.143)&(0.094)\\
    $\mathit{Mention}$&-1.184^{***}&-1.284^{***}&-1.168^{***}&-1.301^{***}\\
    &(0.101)&(0.071)&(0.108)&(0.068)\\
    $\mathit{AcctAge}$&-0.017&-0.040&-0.004&-0.032\\
    &(0.034)&(0.024)&(0.033)&(0.025)\\
    $\mathit{Verified}$&0.604^{***}&0.817^{***}&0.508^{***}&0.797^{***}\\
    &(0.082)&(0.061)&(0.083)&(0.060)\\
    $\mathit{Followers}$&0.160^{***}&0.181^{***}&0.199^{***}&0.224^{***}\\
    &(0.015)&(0.012)&(0.015)&(0.012)\\
    $\mathit{Followees}$&0.037&0.037^{*}&0.042&0.044^{***}\\
    &(0.035)&(0.015)&(0.039)&(0.011)\\
    MonthYear\\
    $2022/08$&0.007&&-0.026&\\
    &(0.131)&&(0.147)&\\
    $2022/09$&0.040&&-0.077&\\
    &(0.144)&&(0.148)&\\
    $2022/10$&-0.056&&-0.181&\\
    &(0.295)&&(0.309)&\\
    $2022/11$&-0.029&-0.038&-0.041&0.111\\
    &(0.342)&(0.233)&(0.350)&(0.213)\\
    $2022/12$&-0.244&-0.296&-0.270&-0.179\\
    &(0.349)&(0.242)&(0.360)&(0.224)\\
    $2023/01$&&-0.785^{**}&&-0.922^{***}\\
    &&(0.258)&&(0.246)\\
    $2023/02$&&-1.186^{***}&&-1.400^{***}\\
    &&(0.263)&&(0.253)\\
    $2023/03$&&-1.229^{***}&&-1.389^{***}\\
    &&(0.261)&&(0.249)\\
    $2023/04$&&-1.055^{***}&&-1.355^{***}\\
    &&(0.271)&&(0.255)\\
    Emotions&$\checkmark$&$\checkmark$&$\checkmark$&$\checkmark$\\
    Topics&$\checkmark$&$\checkmark$&$\checkmark$&$\checkmark$\\
    $\mathit{Intercept}$&7.024^{***}&6.941^{***}&8.816^{***}&8.490^{***}\\
    &(0.151)&(0.166)&(0.163)&(0.155)\\
    Observations&{5,019}&{11,821}&{5,019}&{11,821}\\
    \bottomrule
    \multicolumn{5}{l}{\makecell[l]{Negative binomial regression models with robust standard errors \\ in parentheses. $^*p<0.05$, $^{**}p<0.01$, and $^{***}p<0.001$.}}\\
\end{tabular}}
\end{table}

\clearpage
\begin{table}
\caption{DiD regression results with external controls for retweet count plus, like count plus, and retweets including quotes within F-CRH (treatment) and F-NCRH (control) groups.}
\label{tab:mis_all_did_ext_plus}
\resizebox{0.8\textwidth}{!}{
\begin{tabular}{l*{6}{S}}
    \toprule
    &\multicolumn{2}{c}{Retweet Count Plus}&\multicolumn{2}{c}{Like Count Plus}&\multicolumn{2}{c}{Retweets $+$ Quotes}\\
    \cmidrule(r){2-3} \cmidrule(r){4-5} \cmidrule(r){6-7}
    &{\makecell{(1)\\Pre-US}}&{\makecell{(2)\\US-World}}&{\makecell{(3)\\Pre-US}}&{\makecell{(4)\\US-World}}&{\makecell{(5)\\Pre-US}}&{\makecell{(6)\\US-World}}\\
    \midrule
    $\mathit{FCRH}$&1.174^{***}&0.865^{***}&1.225^{***}&0.949^{***}&1.163^{***}&0.911^{***}\\
    &(0.133)&(0.114)&(0.140)&(0.117)&(0.128)&(0.120)\\
    $\mathit{Vis}$&0.179&0.019&0.120&0.197&0.217&0.053\\
    &(0.300)&(0.122)&(0.310)&(0.135)&(0.288)&(0.123)\\
    $\mathit{FCRH \times Vis}$&-0.430^{**}&-0.053&-0.410^{*}&-0.073&-0.375^{*}&-0.079\\
    &(0.161)&(0.129)&(0.168)&(0.135)&(0.160)&(0.135)\\
    $\mathit{Musk}$&0.096&0.155&0.193&0.211&0.041&0.104\\
    &(0.197)&(0.254)&(0.186)&(0.238)&(0.206)&(0.272)\\
    $\mathit{OldWriter}$&0.029&-0.101&0.034&-0.025&0.019&-0.104\\
    &(0.090)&(0.068)&(0.093)&(0.069)&(0.094)&(0.068)\\
    $\mathit{WordNum}$&0.081^{*}&0.136^{***}&-0.072&-0.010&0.074&0.128^{***}\\
    &(0.037)&(0.030)&(0.041)&(0.032)&(0.038)&(0.030)\\
    $\mathit{Media}$&0.505^{***}&0.740^{***}&0.454^{***}&0.595^{***}&0.583^{***}&0.790^{***}\\
    &(0.076)&(0.057)&(0.082)&(0.059)&(0.079)&(0.057)\\
    $\mathit{Hashtag}$&-0.863^{***}&-0.696^{***}&-1.057^{***}&-0.890^{***}&-0.828^{***}&-0.714^{***}\\
    &(0.148)&(0.089)&(0.143)&(0.093)&(0.155)&(0.090)\\
    $\mathit{Mention}$&-1.176^{***}&-1.252^{***}&-1.167^{***}&-1.294^{***}&-1.145^{***}&-1.258^{***}\\
    &(0.100)&(0.068)&(0.107)&(0.067)&(0.104)&(0.070)\\
    $\mathit{AcctAge}$&-0.017&-0.037&-0.004&-0.031&-0.015&-0.040\\
    &(0.034)&(0.024)&(0.033)&(0.024)&(0.035)&(0.024)\\
    $\mathit{Verified}$&0.602^{***}&0.814^{***}&0.508^{***}&0.796^{***}&0.664^{***}&0.843^{***}\\
    &(0.081)&(0.060)&(0.082)&(0.060)&(0.083)&(0.060)\\
    $\mathit{Followers}$&0.160^{***}&0.180^{***}&0.199^{***}&0.224^{***}&0.161^{***}&0.188^{***}\\
    &(0.015)&(0.012)&(0.015)&(0.012)&(0.015)&(0.012)\\
    $\mathit{Followees}$&0.035&0.036^{*}&0.042&0.044^{***}&0.028&0.036^{*}\\
    &(0.035)&(0.015)&(0.039)&(0.011)&(0.034)&(0.015)\\
    MonthYear\\
    $2022/08$&0.007&&-0.026&&0.001&\\
    &(0.130)&&(0.147)&&(0.130)&\\
    $2022/09$&0.043&&-0.076&&0.004&\\
    &(0.143)&&(0.148)&&(0.141)&\\
    $2022/10$&-0.051&&-0.179&&-0.078&\\
    &(0.292)&&(0.308)&&(0.281)&\\
    $2022/11$&-0.018&-0.027&-0.039&0.112&-0.089&-0.073\\
    &(0.338)&(0.226)&(0.349)&(0.212)&(0.334)&(0.247)\\
    $2022/12$&-0.229&-0.282&-0.267&-0.177&-0.334&-0.358\\
    &(0.345)&(0.235)&(0.359)&(0.223)&(0.340)&(0.254)\\
    $2023/01$&&-0.755^{**}&&-0.916^{***}&&-0.845^{**}\\
    &&(0.251)&&(0.245)&&(0.270)\\
    $2023/02$&&-1.148^{***}&&-1.392^{***}&&-1.281^{***}\\
    &&(0.256)&&(0.252)&&(0.273)\\
    $2023/03$&&-1.190^{***}&&-1.381^{***}&&-1.304^{***}\\
    &&(0.254)&&(0.248)&&(0.272)\\
    $2023/04$&&-1.018^{***}&&-1.347^{***}&&-1.157^{***}\\
    &&(0.263)&&(0.254)&&(0.281)\\
    Emotions&$\checkmark$&$\checkmark$&$\checkmark$&$\checkmark$&$\checkmark$&$\checkmark$\\
    Topics&$\checkmark$&$\checkmark$&$\checkmark$&$\checkmark$&$\checkmark$&$\checkmark$\\
    $\mathit{Intercept}$&7.030^{***}&6.952^{***}&8.817^{***}&8.493^{***}&7.200^{***}&7.160^{***}\\
    &(0.150)&(0.165)&(0.163)&(0.154)&(0.148)&(0.162)\\
    Observations&{5,019}&{11,821}&{5,019}&{11,821}&{5,019}&{11,821}\\
    \bottomrule
    \multicolumn{7}{l}{\makecell[l]{Negative binomial regression models with robust standard errors \\ in parentheses. $^*p<0.05$, $^{**}p<0.01$, and $^{***}p<0.001$.}}\\
\end{tabular}}
\end{table}

\clearpage
\begin{table}
\caption{DiD regression results with external controls for retweet count and like count within F-CRH (treatment) and F-NCRH (control) matched groups during different periods.}
\label{tab:mis_all_did_ext_matched}
\resizebox{0.6\textwidth}{!}{
\begin{tabular}{l*{4}{S}}
    \toprule
    &\multicolumn{2}{c}{Retweet Count}&\multicolumn{2}{c}{Like Count}\\
    \cmidrule(r){2-3} \cmidrule(r){4-5}
    &{\makecell{(1)\\Pre-US}}&{\makecell{(2)\\US-World}}&{\makecell{(3)\\Pre-US}}&{\makecell{(4)\\US-World}}\\
    \midrule
    $\mathit{FCRH}$&1.177^{***}&0.830^{***}&1.220^{***}&0.896^{***}\\
    &(0.136)&(0.109)&(0.141)&(0.108)\\
    $\mathit{Vis}$&0.076&0.053&-0.012&0.163\\
    &(0.312)&(0.146)&(0.326)&(0.165)\\
    $\mathit{FCRH \times Vis}$&-0.454^{*}&0.030&-0.390^{*}&0.120\\
    &(0.182)&(0.148)&(0.186)&(0.153)\\
    $\mathit{Musk}$&0.236&0.149&0.257&0.217\\
    &(0.261)&(0.238)&(0.245)&(0.216)\\
    $\mathit{OldWriter}$&-0.019&-0.115&0.045&-0.071\\
    &(0.120)&(0.072)&(0.123)&(0.073)\\
    $\mathit{WordNum}$&0.087&0.120^{**}&-0.053&-0.003\\
    &(0.045)&(0.039)&(0.048)&(0.041)\\
    $\mathit{Media}$&0.502^{***}&0.699^{***}&0.451^{***}&0.577^{***}\\
    &(0.090)&(0.076)&(0.099)&(0.074)\\
    $\mathit{Hashtag}$&-0.898^{***}&-0.835^{***}&-1.085^{***}&-1.045^{***}\\
    &(0.155)&(0.138)&(0.152)&(0.131)\\
    $\mathit{Mention}$&-1.198^{***}&-1.241^{***}&-1.197^{***}&-1.179^{***}\\
    &(0.123)&(0.104)&(0.132)&(0.101)\\
    $\mathit{AcctAge}$&0.004&-0.089^{**}&-0.001&-0.076^{*}\\
    &(0.041)&(0.032)&(0.042)&(0.032)\\
    $\mathit{Verified}$&0.627^{***}&0.863^{***}&0.515^{***}&0.796^{***}\\
    &(0.094)&(0.081)&(0.097)&(0.078)\\
    $\mathit{Followers}$&0.170^{***}&0.183^{***}&0.232^{***}&0.219^{***}\\
    &(0.031)&(0.014)&(0.038)&(0.013)\\
    $\mathit{Followees}$&0.043&0.024&0.052&0.046^{*}\\
    &(0.043)&(0.022)&(0.042)&(0.021)\\
    MonthYear\\
    $2022/08$&-0.014&&-0.047&\\
    &(0.130)&&(0.144)&\\
    $2022/09$&0.042&&-0.077&\\
    &(0.146)&&(0.148)&\\
    $2022/10$&-0.032&&-0.122&\\
    &(0.298)&&(0.321)&\\
    $2022/11$&-0.118&-0.023&0.010&0.107\\
    &(0.392)&(0.209)&(0.410)&(0.187)\\
    $2022/12$&-0.344&-0.291&-0.198&-0.187\\
    &(0.404)&(0.218)&(0.425)&(0.200)\\
    $2023/01$&&-0.907^{***}&&-0.954^{***}\\
    &&(0.259)&&(0.253)\\
    $2023/02$&&-1.226^{***}&&-1.441^{***}\\
    &&(0.259)&&(0.255)\\
    $2023/03$&&-1.317^{***}&&-1.447^{***}\\
    &&(0.257)&&(0.251)\\
    $2023/04$&&-1.205^{***}&&-1.487^{***}\\
    &&(0.295)&&(0.277)\\
    Emotions&$\checkmark$&$\checkmark$&$\checkmark$&$\checkmark$\\
    Topics&$\checkmark$&$\checkmark$&$\checkmark$&$\checkmark$\\
    $\mathit{Intercept}$&7.123^{***}&6.959^{***}&8.875^{***}&8.567^{***}\\
    &(0.177)&(0.184)&(0.187)&(0.165)\\
    Observations&{3,814}&{6,021}&{3,814}&{6,021}\\
    \bottomrule
    \multicolumn{5}{l}{\makecell[l]{Negative binomial regression models with robust standard errors \\ in parentheses. $^*p<0.05$, $^{**}p<0.01$, and $^{***}p<0.001$.}}\\
\end{tabular}}
\end{table}

\clearpage
\begin{table}
\caption{DiD regression results with external controls for retweet count and like count within F-CRH (treatment) and OCN (control) groups.}
\label{tab:mis_out_did}
\resizebox{0.6\textwidth}{!}{
\begin{tabular}{l*{4}{S}}
    \toprule
    &\multicolumn{2}{c}{Retweet Count}&\multicolumn{2}{c}{Like Count}\\
    \cmidrule(r){2-3} \cmidrule(r){4-5}
    &{\makecell{(1)\\Pre-US}}&{\makecell{(2)\\US-World}}&{\makecell{(3)\\Pre-US}}&{\makecell{(4)\\US-World}}\\
    \midrule
    $\mathit{FCRH}$&2.835^{***}&2.464^{***}&2.623^{***}&2.460^{***}\\
    &(0.315)&(0.275)&(0.284)&(0.254)\\
    $\mathit{Vis}$&0.220&-0.151&0.144&0.102\\
    &(0.552)&(0.351)&(0.532)&(0.361)\\
    $\mathit{FCRH \times Vis}$&-0.669&-0.121&-0.524&-0.401\\
    &(0.422)&(0.305)&(0.429)&(0.290)\\
    $\mathit{Musk}$&-0.133&-0.200&-0.230&-0.193\\
    &(0.570)&(0.400)&(0.554)&(0.416)\\
    $\mathit{WordNum}$&-0.098&-0.091&-0.259&-0.194^{**}\\
    &(0.134)&(0.061)&(0.136)&(0.067)\\
    $\mathit{Media}$&0.284&0.327^{**}&0.522^{*}&0.402^{***}\\
    &(0.259)&(0.115)&(0.246)&(0.118)\\
    $\mathit{Hashtag}$&-0.802^{*}&-1.423^{***}&-0.904^{*}&-1.528^{***}\\
    &(0.403)&(0.169)&(0.373)&(0.179)\\
    $\mathit{Mention}$&-0.880^{*}&-0.487^{**}&-0.667&-0.571^{**}\\
    &(0.352)&(0.177)&(0.373)&(0.181)\\
    $\mathit{AcctAge}$&0.371^{**}&-0.144^{**}&0.326^{**}&-0.174^{**}\\
    &(0.122)&(0.054)&(0.122)&(0.055)\\
    $\mathit{Verified}$&0.009&0.572^{***}&-0.075&0.687^{***}\\
    &(0.241)&(0.135)&(0.225)&(0.137)\\
    $\mathit{Followers}$&0.362^{***}&0.143^{***}&0.396^{***}&0.156^{***}\\
    &(0.092)&(0.022)&(0.106)&(0.024)\\
    $\mathit{Followees}$&0.011&-0.044&0.113&-0.054\\
    &(0.116)&(0.032)&(0.173)&(0.037)\\
    MonthYear\\
    $2022/08$&0.132&&-0.184&\\
    &(0.331)&&(0.318)&\\
    $2022/09$&-0.049&&-0.032&\\
    &(0.286)&&(0.292)&\\
    $2022/10$&-0.165&&-0.376&\\
    &(0.402)&&(0.371)&\\
    $2022/11$&0.481&0.287&0.574&0.468\\
    &(0.674)&(0.371)&(0.626)&(0.392)\\
    $2022/12$&0.445&0.057&0.590&0.110\\
    &(0.752)&(0.401)&(0.694)&(0.421)\\
    $2023/01$&&-0.376&&-0.351\\
    &&(0.471)&&(0.503)\\
    $2023/02$&&0.043&&-0.001\\
    &&(0.472)&&(0.501)\\
    $2023/03$&&-0.127&&-0.088\\
    &&(0.458)&&(0.493)\\
    $2023/04$&&-0.586&&-0.341\\
    &&(0.498)&&(0.536)\\
    Emotions&$\checkmark$&$\checkmark$&$\checkmark$&$\checkmark$\\
    Topics&$\checkmark$&$\checkmark$&$\checkmark$&$\checkmark$\\
    $\mathit{Intercept}$&5.952^{***}&5.441^{***}&7.877^{***}&7.096^{***}\\
    &(0.380)&(0.352)&(0.351)&(0.319)\\
    Observations&{538}&{1,874}&{538}&{1,874}\\
    \bottomrule
    \multicolumn{5}{l}{\makecell[l]{Negative binomial regression models with robust standard errors \\ in parentheses. $^*p<0.05$, $^{**}p<0.01$, and $^{***}p<0.001$.}}\\
\end{tabular}}
\end{table}

\clearpage
\begin{table}
\caption{Regression results for retweet count parallel test between F-CRH and F-CNRH groups.}
\label{tab:parallel_mis_did_retweet}
\begin{tabular}{lcccccc}
\toprule
Variable&Coef.&Robust Std. Err.&z&P>|z|&\multicolumn{2}{c}{95\%CI}\\
\midrule
$\mathit{FCRH}$&1.402&0.251&5.587&0.000&0.910&1.893\\
$\mathit{FCRH \times Pre4}$&0.537&0.360&1.493&0.136&-0.168&1.242\\
$\mathit{FCRH \times Pre3}$&0.612&0.363&1.688&0.091&-0.099&1.322\\
$\mathit{FCRH \times Pre2}$&0.157&0.285&0.550&0.582&-0.402&0.716\\
$\mathit{FCRH \times Pre1}$&-0.167&0.285&-0.586&0.558&-0.726&0.392\\
$\mathit{FCRH \times VisUS}$&-0.576&0.277&-2.083&0.037&-1.119&-0.034\\
$\mathit{FCRH \times VisWorld}$&-0.708&0.257&-2.752&0.006&-1.212&-0.204\\
$\mathit{WordNum}$&0.119&0.022&5.365&0.000&0.076&0.163\\
$\mathit{Media}$&0.496&0.043&11.632&0.000&0.413&0.580\\
$\mathit{Hashtag}$&-0.469&0.086&-5.436&0.000&-0.638&-0.300\\
$\mathit{Mention}$&-1.271&0.058&-21.776&0.000&-1.386&-1.157\\
$\mathit{AcctAge}$&-0.066&0.020&-3.355&0.001&-0.104&-0.027\\
$\mathit{Verified}$&0.695&0.044&15.772&0.000&0.609&0.781\\
$\mathit{Followers}$&0.189&0.012&16.027&0.000&0.166&0.212\\
$\mathit{Followees}$&0.063&0.017&3.741&0.000&0.030&0.095\\
Emotions&\multicolumn{6}{c}{$\checkmark$}\\
Topics&\multicolumn{6}{c}{$\checkmark$}\\
Periods&\multicolumn{6}{c}{$\checkmark$}\\
$\mathit{Intercept}$&6.088&0.069&88.591&0.000&5.953&6.222\\
Observations&\multicolumn{6}{c}{24,018}\\
\bottomrule
\multicolumn{7}{l}{\makecell[l]{Negative binomial regression model with robust standard errors.}}\\
\end{tabular}
\end{table}

\begin{table}
\caption{Regression results for like count parallel test between F-CRH and F-CNRH groups.}
\label{tab:parallel_mis_did_like}
\begin{tabular}{lcccccc}
\toprule
Variable&Coef.&Robust Std. Err.&z&P>|z|&\multicolumn{2}{c}{95\%CI}\\
\midrule
$\mathit{FCRH}$&1.670&0.293&5.694&0.000&1.095&2.245\\
$\mathit{FCRH \times Pre4}$&0.302&0.372&0.811&0.417&-0.428&1.031\\
$\mathit{FCRH \times Pre3}$&0.596&0.387&1.543&0.123&-0.161&1.354\\
$\mathit{FCRH \times Pre2}$&0.137&0.323&0.424&0.671&-0.495&0.769\\
$\mathit{FCRH \times Pre1}$&-0.371&0.327&-1.137&0.256&-1.012&0.269\\
$\mathit{FCRH \times VisUS}$&-0.740&0.315&-2.352&0.019&-1.357&-0.124\\
$\mathit{FCRH \times VisWorld}$&-0.876&0.300&-2.921&0.003&-1.463&-0.288\\
$\mathit{WordNum}$&-0.036&0.023&-1.560&0.119&-0.081&0.009\\
$\mathit{Media}$&0.334&0.046&7.208&0.000&0.243&0.425\\
$\mathit{Hashtag}$&-0.697&0.075&-9.261&0.000&-0.845&-0.550\\
$\mathit{Mention}$&-1.289&0.056&-22.974&0.000&-1.399&-1.179\\
$\mathit{AcctAge}$&-0.057&0.021&-2.711&0.007&-0.098&-0.016\\
$\mathit{Verified}$&0.611&0.047&12.964&0.000&0.518&0.703\\
$\mathit{Followers}$&0.239&0.011&21.088&0.000&0.217&0.261\\
$\mathit{Followees}$&0.054&0.013&4.135&0.000&0.028&0.080\\
Emotions&\multicolumn{6}{c}{$\checkmark$}\\
Topics&\multicolumn{6}{c}{$\checkmark$}\\
Periods&\multicolumn{6}{c}{$\checkmark$}\\
$\mathit{Intercept}$&7.656&0.071&108.502&0.000&7.518&7.794\\
Observations&\multicolumn{6}{c}{24,018}\\
\bottomrule
\multicolumn{7}{l}{\makecell[l]{Negative binomial regression model with robust standard errors.}}\\
\end{tabular}
\end{table}

\clearpage
\begin{table}
\caption{Regression results for retweet count parallel test between T-NMR and F-CNRH groups.}
\label{tab:parallel_mis_did_retweet_placebo}
\begin{tabular}{lcccccc}
\toprule
Variable&Coef.&Robust Std. Err.&z&P>|z|&\multicolumn{2}{c}{95\%CI}\\
\midrule
$\mathit{TNMR}$&1.841&0.120&15.313&0.000&1.605&2.076\\
$\mathit{TNMR \times Pre4}$&-0.131&0.200&-0.658&0.511&-0.523&0.260\\
$\mathit{TNMR \times Pre3}$&0.031&0.256&0.121&0.904&-0.471&0.533\\
$\mathit{TNMR \times Pre2}$&-0.086&0.177&-0.488&0.625&-0.433&0.260\\
$\mathit{TNMR \times Pre1}$&-0.243&0.230&-1.058&0.290&-0.693&0.207\\
$\mathit{TNMR \times VisUS}$&-0.796&0.149&-5.337&0.000&-1.088&-0.504\\
$\mathit{TNMR \times VisWorld}$&-0.742&0.133&-5.601&0.000&-1.002&-0.483\\
$\mathit{WordNum}$&0.109&0.021&5.265&0.000&0.068&0.149\\
$\mathit{Media}$&0.466&0.040&11.524&0.000&0.387&0.545\\
$\mathit{Hashtag}$&-0.421&0.084&-5.029&0.000&-0.585&-0.257\\
$\mathit{Mention}$&-1.184&0.054&-22.015&0.000&-1.289&-1.078\\
$\mathit{AcctAge}$&-0.079&0.018&-4.328&0.000&-0.115&-0.043\\
$\mathit{Verified}$&0.714&0.043&16.560&0.000&0.630&0.799\\
$\mathit{Followers}$&0.249&0.014&17.797&0.000&0.221&0.276\\
$\mathit{Followees}$&0.051&0.017&2.966&0.003&0.017&0.084\\
Emotions&\multicolumn{6}{c}{$\checkmark$}\\
Topics&\multicolumn{6}{c}{$\checkmark$}\\
Periods&\multicolumn{6}{c}{$\checkmark$}\\
$\mathit{Intercept}$&6.083&0.067&91.447&0.000&5.953&6.214\\
Observations&\multicolumn{6}{c}{26,016}\\
\bottomrule
\multicolumn{7}{l}{\makecell[l]{Negative binomial regression model with robust standard errors.}}\\
\end{tabular}
\end{table}

\begin{table}
\caption{Regression results for like count parallel test between T-NMR and F-CNRH groups.}
\label{tab:parallel_mis_did_like_placebo}
\begin{tabular}{lcccccc}
\toprule
Variable&Coef.&Robust Std. Err.&z&P>|z|&\multicolumn{2}{c}{95\%CI}\\
\midrule
$\mathit{TNMR}$&1.884&0.117&16.143&0.000&1.655&2.113\\
$\mathit{TNMR \times Pre4}$&-0.043&0.207&-0.206&0.837&-0.448&0.363\\
$\mathit{TNMR \times Pre3}$&-0.020&0.242&-0.082&0.935&-0.493&0.454\\
$\mathit{TNMR \times Pre2}$&-0.117&0.170&-0.685&0.494&-0.451&0.217\\
$\mathit{TNMR \times Pre1}$&-0.271&0.211&-1.285&0.199&-0.686&0.143\\
$\mathit{TNMR \times VisUS}$&-0.781&0.147&-5.313&0.000&-1.069&-0.493\\
$\mathit{TNMR \times VisWorld}$&-0.709&0.131&-5.433&0.000&-0.965&-0.453\\
$\mathit{Media}$&0.270&0.043&6.252&0.000&0.186&0.355\\
$\mathit{Hashtag}$&-0.635&0.072&-8.828&0.000&-0.775&-0.494\\
$\mathit{Mention}$&-1.195&0.052&-23.108&0.000&-1.297&-1.094\\
$\mathit{AcctAge}$&-0.069&0.020&-3.519&0.000&-0.108&-0.031\\
$\mathit{Verified}$&0.609&0.046&13.281&0.000&0.520&0.699\\
$\mathit{Followers}$&0.318&0.013&23.661&0.000&0.292&0.344\\
$\mathit{Followees}$&0.043&0.013&3.242&0.001&0.017&0.069\\
Emotions&\multicolumn{6}{c}{$\checkmark$}\\
Topics&\multicolumn{6}{c}{$\checkmark$}\\
Periods&\multicolumn{6}{c}{$\checkmark$}\\
$\mathit{Intercept}$&7.670&0.068&112.803&0.000&7.537&7.803\\
Observations&\multicolumn{6}{c}{26,016}\\
\bottomrule
\multicolumn{7}{l}{\makecell[l]{Negative binomial regression model with robust standard errors.}}\\
\end{tabular}
\end{table}

\clearpage
\begin{table}
\caption{Regression results for retweet count parallel test between F-CRH and T-NMR groups.}
\label{tab:parallel_tf_retweet}
\begin{tabular}{lcccccc}
\toprule
Variable&Coef.&Robust Std. Err.&z&P>|z|&\multicolumn{2}{c}{95\%CI}\\
\midrule
$\mathit{FCRH}$&-0.637&0.220&-2.895&0.004&-1.068&-0.206\\
$\mathit{FCRH \times Pre4}$&0.579&0.335&1.730&0.084&-0.077&1.235\\
$\mathit{FCRH \times Pre3}$&0.307&0.422&0.728&0.467&-0.520&1.135\\
$\mathit{FCRH \times Pre2}$&0.169&0.262&0.642&0.521&-0.346&0.683\\
$\mathit{FCRH \times Pre1}$&0.318&0.285&1.119&0.263&-0.239&0.876\\
$\mathit{FCRH \times VisUS}$&0.173&0.244&0.707&0.479&-0.306&0.651\\
$\mathit{FCRH \times VisWorld}$&0.223&0.230&0.970&0.332&-0.227&0.673\\
$\mathit{WordNum}$&0.021&0.026&0.821&0.412&-0.029&0.071\\
$\mathit{Media}$&0.245&0.051&4.827&0.000&0.145&0.344\\
$\mathit{Hashtag}$&-0.452&0.096&-4.727&0.000&-0.639&-0.265\\
$\mathit{Mention}$&-0.587&0.065&-9.049&0.000&-0.714&-0.460\\
$\mathit{AcctAge}$&-0.074&0.023&-3.161&0.002&-0.119&-0.028\\
$\mathit{Verified}$&0.082&0.054&1.517&0.129&-0.024&0.187\\
$\mathit{Followers}$&0.291&0.020&14.574&0.000&0.252&0.330\\
$\mathit{Followees}$&-0.016&0.012&-1.334&0.182&-0.038&0.007\\
Emotions&\multicolumn{6}{c}{$\checkmark$}\\
Topics&\multicolumn{6}{c}{$\checkmark$}\\
Periods&\multicolumn{6}{c}{$\checkmark$}\\
$\mathit{Intercept}$&8.479&0.113&75.182&0.000&8.258&8.701\\
Observations&\multicolumn{6}{c}{8,499}\\
\bottomrule
\multicolumn{7}{l}{\makecell[l]{Negative binomial regression model with robust standard errors.}}\\
\end{tabular}
\end{table}

\clearpage
\begin{table}
\caption{Regression results for like count parallel test between F-CRH and T-NMR groups.}
\label{tab:parallel_tf_like}
\begin{tabular}{lcccccc}
\toprule
Variable&Coef.&Robust Std. Err.&z&P>|z|&\multicolumn{2}{c}{95\%CI}\\
\midrule
$\mathit{FCRH}$&-0.527&0.251&-2.102&0.036&-1.019&-0.036\\
$\mathit{FCRH \times Pre4}$&0.462&0.354&1.303&0.193&-0.233&1.156\\
$\mathit{FCRH \times Pre3}$&0.570&0.412&1.383&0.167&-0.237&1.377\\
$\mathit{FCRH \times Pre2}$&0.277&0.291&0.952&0.341&-0.293&0.847\\
$\mathit{FCRH \times Pre1}$&0.220&0.301&0.730&0.465&-0.370&0.809\\
$\mathit{FCRH \times VisUS}$&0.153&0.273&0.559&0.576&-0.383&0.689\\
$\mathit{FCRH \times VisWorld}$&0.155&0.263&0.591&0.554&-0.359&0.670\\
$\mathit{WordNum}$&-0.149&0.029&-5.099&0.000&-0.206&-0.092\\
$\mathit{Media}$&0.097&0.055&1.750&0.080&-0.012&0.206\\
$\mathit{Hashtag}$&-0.586&0.101&-5.816&0.000&-0.783&-0.388\\
$\mathit{Mention}$&-0.636&0.066&-9.657&0.000&-0.765&-0.507\\
$\mathit{AcctAge}$&-0.072&0.025&-2.940&0.003&-0.121&-0.024\\
$\mathit{Verified}$&0.051&0.056&0.909&0.363&-0.059&0.161\\
$\mathit{Followers}$&0.336&0.018&19.047&0.000&0.302&0.371\\
$\mathit{Followees}$&0.004&0.013&0.313&0.754&-0.021&0.029\\
Emotions&\multicolumn{6}{c}{$\checkmark$}\\
Topics&\multicolumn{6}{c}{$\checkmark$}\\
Periods&\multicolumn{6}{c}{$\checkmark$}\\
$\mathit{Intercept}$&10.111&0.113&89.632&0.000&9.890&10.332\\
Observations&\multicolumn{6}{c}{8,499}\\
\bottomrule
\multicolumn{7}{l}{\makecell[l]{Negative binomial regression model with robust standard errors.}}\\
\end{tabular}
\end{table}

\end{document}